%% file: dec_nlo.tex
\documentclass[12pt]{article}
\usepackage{a4wide,epsfig,amsmath,amssymb,cite}
\usepackage{feynarts}

\newcommand{\lM}{}
\newcommand{\lt}{}
\newcommand{\lgl}{}
\newcommand{\ltM}{}

\newcommand{\asfull}{\alpha_s^{\rm (full)}}
\newcommand{\asA}{\alpha_s^{(\tilde{t},\tilde{g},6)}}
\newcommand{\asB}{\alpha_s^{(\tilde{g},6)}}
\newcommand{\asC}{\alpha_s^{(6)}}

\newcommand{\asDRbar}{\alpha_s^{\overline{\rm DR}}}
\newcommand{\asMSbar}{\alpha_s^{\overline{\rm MS}}}

\begin{document}

\title{\vskip-3cm{\baselineskip14pt
    \begin{flushleft}
      \normalsize SFB/CPP-05-41 \\
      \normalsize TTP/05-14 \\
      \normalsize hep-ph/0509048\\
      \normalsize September 2005\\ 
  \end{flushleft}}
  \vskip1.5cm
  Two-loop matching coefficients for the strong coupling in the MSSM
}
\author{\small R. Harlander, L. Mihaila, M. Steinhauser\\
  {\small\it Institut f{\"u}r Theoretische Teilchenphysik,
    Universit{\"a}t Karlsruhe,}\\
  {\small\it 76128 Karlsruhe, Germany}\\
}

\date{}

\maketitle

\thispagestyle{empty}

\begin{abstract}
  When relating the strong coupling $\alpha_s$,
  measured at the scale of the $Z$ boson mass, to
  its numerical value at some higher energy, for example
  the scale of Grand Unification, it is important 
  to include higher order corrections both in the running of $\alpha_s$ and
  the decoupling of the heavy particles.
  We compute the two-loop matching coefficients 
  for $\alpha_s$ within the 
  Minimal Supersymmetric Standard Model (MSSM) which are necessary for a
  consistent three-loop evolution of the strong coupling constant.
  Different scenarios for the hierarchy of the
  supersymmetric scales are considered and the 
  numerical effects are discussed. We find that the three-loop effects
  can be as large as and sometimes even larger than the uncertainty 
  induced by the current experimental accuracy of $\alpha_s(M_Z)$.
\medskip

\noindent
PACS: 12.10.Kt, 12.38.-t, 12.38.Bx, 12.60.Jv 

\end{abstract}

\newpage

\section{Introduction}

In spite of describing all current experimental precision data extremely
well, it is commonly believed that the Standard Model (SM) is not the
ultimate theory of particle physics.  This is due to deficiencies like
the so-called fine tuning problem which arises from quadratic
divergences combined with the large difference between the electro-weak
and the Planck scale.

A popular extension of the SM that cures this problem is the so-called
Minimal Supersymmetric Standard Model (MSSM). In addition, one of its
most attractive features is the possibility of unification of the
strong, the weak, and the electro-magnetic coupling constants. While
having very different numerical values at the electro-weak scale, they
may have a common intersection point at a scale $\mu_{\rm GUT} \sim
10^{16}$~GeV~\cite{Dimopoulos:1981yj,Ibanez:1981yh,Amaldi:1991cn,
Langacker:1991an,Ellis:1990wk,Blair:2002pg} which is therefore
considered to be the typical scale of a Grand Unified Theory (GUT).
In the minimal SM, such an intersection can not be achieved.

The relation of the coupling constants at different energy scales is
governed by the renormalization group running, accompanied by the
appropriate decoupling of heavy particles, to be described in more
detail in Section~\ref{sec::decoupling}.  The most recent studies of the
coupling constants at the GUT scale as evolved from their experimental
values at the weak scale have been performed in the context of the SUSY
Parameter Analysis project (SPA)~\cite{SPA}. It is the purpose of this
paper to determine the precise evolution of the strong coupling
$\alpha_s$.  In particular, we evaluate the two-loop decoupling
relations within the MSSM. In combination with the known beta function,
this allows us to determine $\alpha_s(\mu_{\rm GUT})$ at three-loop
accuracy, and to estimate its theoretical uncertainty.

The outline of this paper is as follows: in
Sections~\ref{sec::decoupling} and \ref{sec::tech} we review the
theoretical framework and introduce our notation.  In
Section~\ref{sec::renconst} we quote the renormalization constants and
the $\beta$ function in the $\overline{\rm DR}$ scheme within the MSSM.
Our results for the decoupling relations are given in
Section~\ref{sec::decconst}, and the phenomenological implications are
discussed in Section~\ref{sec::num}.  Our findings are summarized in
Section~\ref{sec::concl}.

\section{Decoupling of heavy particles}\label{sec::decoupling}

In the framework of QCD and its supersymmetric extension, it is
convenient to use a mass-independent renormalization scheme such as
$\overline {\rm MS}$ or $\overline {\rm DR}$. Thus, by construction, the
beta function governing the running of $\alpha_s$ is independent of the
particle masses. It only depends on the particle content of the
underlying theory, i.e., the number of active quarks, squarks and
gluinos, denoted by $n_f$, $n_s$, and $n_{\tilde{g}}$ in what follows (a
more precise definition of these labels is given subsequent to
Eq.~(\ref{eq::renconst_res})).

The price for this simplicity is that, in contrast to momentum
subtraction schemes, the Appelquist-Carrazone decoupling
theorem~\cite{Appelquist:1974tg} does not hold. This is well-known in
pure QCD where, at an energy scale $\mu$, Green functions involving
heavy quarks of mass $m_h$ depend logarithmically on $\mu/m_h$. In order
to avoid potentially large logarithms for $\mu\ll m_h$ one has to
decouple the heavy quark from the theory~\cite{Bernreuther:1981sg}.  As
a consequence, one obtains different coupling constants in the various
energy regimes.  They are connected by the so-called decoupling or
matching relations.  This is usually indicated by labeling $\alpha_s$
with a superscript which determines the number of active particles that
contribute to its running.  Considered as a function of $\mu$,
$\alpha_s$ shows a kink at one-loop and even a jump at higher orders
when changing from one energy regime to the other.

The precise value where the heavy degrees of freedom are integrated out
is not fixed by theory. Since the explicit dependence on this matching
scale is logarithmic, it is natural to choose $\mu\approx m_h$.  The
dependence of theoretical predictions on the variation of the matching
scale around this intuitive value can be used as an estimate of the
theoretical uncertainty. Reduction of this uncertainty can only be
achieved by means of higher order calculations.

Within QCD, the decoupling relations have been computed up to the
three-loop order~\cite{Chetyrkin:1997un}. In the MSSM, only the one-loop
relation for $\alpha_s$ is known~\cite{Hall:1980kf,Harlander:2004tp}.  It
is the purpose of this paper to extend this calculation to two loops,
such that a consistent three-loop evolution of $\alpha_s$ to the GUT
scale can be performed.  The matching coefficients that relate different
energy regimes (with different values of $n_s$, $n_f$, and $n_{\tilde
g}$) involve the masses of the heavy particles. Whereas at one-loop
order the dependence is only logarithmic, it involves much more
complicated functions at two-loop order.  Furthermore, it is important
whether the heavy particles have approximately similar masses and can
thus be integrated out simultaneously, or whether the mass differences
are big and the decoupling has to be performed step-by-step.

In this paper we will consider different scenarios concerning the
hierarchy among the relevant particle masses.  They are defined by the
following conditions:
\begin{itemize}
\item[(A)]
  $m_{\tilde{u}}$, \ldots, $m_{\tilde{b}}$ 
  $\gg$ 
  $m_{\tilde{t}}$, $m_{\tilde{g}}$, $m_t$
  $\gg m_b$
\item[(B)]
  $m_{\tilde{u}}$, \ldots, $m_{\tilde{b}}$, $m_{\tilde{t}}$ 
  $\gg$ 
  $m_{\tilde{g}}$, $m_t$
  $\gg m_b$
\item[(C)]
  $m_{\tilde{u}}$, \ldots, $m_{\tilde{b}}$, $m_{\tilde{t}}$, $m_{\tilde{g}}$
  $\gg$ 
  $m_t$
  $\gg m_b$
\item[(D)]
  $m_{\tilde{u}}$, \ldots, $m_{\tilde{b}}$, $m_{\tilde{t}}$, $m_{\tilde{g}}$,
  $m_t$
  $\gg m_b\,,$
\end{itemize}  
where $m_{\tilde{u}}$, \ldots, $m_{\tilde{t}}$ are the squark masses,
$m_{\tilde{g}}$ is the gluino mass and $m_t$, $m_b$ are the top and the
bottom quark mass, respectively.\\
In all cases it is understood that masses which are separated by a comma
are of the same order of magnitude.  Note that case~(B) corresponds to
so-called split SUSY which has been receiving much
attention~\cite{Arkani-Hamed:2004fb,Giudice:2004tc} recently.
Furthermore, in case~(C) all supersymmetric
masses are considered as heavy with respect to the SM ones.  Squark
masses are always taken to be larger than the top quark mass.

If all occurring mass scales are different, the resulting formulae are
quite complicated and unhandy.  For the purpose of this paper, we thus
restrict ourselves to the simplified scenarios where either all masses
are identified (if possible), or where an expansion in $m_t/M$ can be
performed, with $M\in\{m_{\tilde u},\ldots,m_{\tilde t},m_{\tilde g}\}$.
Further details on our approximations will be given in
Sections~\ref{sec::tech} and~\ref{sec::decconst}.

\section{\label{sec::tech}Technicalities}

We follow the framework defined in
Ref.~\cite{Chetyrkin:1997sg,Chetyrkin:1997un} (see also
Ref.~\cite{Steinhauser:2002rq}) which reduces the calculation of the
$n$-loop decoupling relation to $n$-loop vacuum diagrams.
The bare decoupling (or matching) coefficients $\zeta^0_i$, $\tilde
\zeta^0_i$ are introduced as
\begin{eqnarray}
  g_s^{0\prime}&=&\zeta_g^0 g_s^0\,,
  \nonumber\\
  G_\mu^{0\prime,a}&=&\sqrt{\zeta_3^0}G_\mu^{0,a}\,,
  \nonumber\\
  c^{0\prime,a}&=&\sqrt{\tilde\zeta_3^0}c^{0,a}\,,
  \nonumber\\
  \Gamma_{ccg}^{0\prime} &=& \tilde\zeta_1^0 \Gamma_{ccg}^0\,,
\end{eqnarray}
where $g_s=\sqrt{4\pi\alpha_s}$ is the QCD gauge coupling, $G_\mu^a$ is
the gluon field, $c^a$ is the Faddeev-Popov ghost field, and
$\Gamma_{ccg}$ is the ghost-gluon vertex.  The prime generically denotes
the quantities in the effective theory, where the heavy particles have
been integrated out and the bare objects are marked by the superscript 
``0''.  According to Ref.~\cite{Chetyrkin:1997un},
$\zeta_3^0$, $\tilde\zeta_3^0$ and $\tilde\zeta_1^0$ are obtained from
the gluon propagator, the ghost propagator, and the ghost-gluon vertex,
all evaluated at vanishing external momenta.  $\zeta_g^0$ is then
determined from the relation
\begin{eqnarray}
  \zeta_g^0 &=&
  \frac{\tilde\zeta_1^0}{\tilde\zeta_3^0\sqrt{\zeta_3^0}}
  \,.
\end{eqnarray}

The renormalized decoupling coefficient is obtained with the help of the
renormalization constants defined through
\begin{eqnarray}
  g_s^0   &=& \mu^{2\epsilon}Z_g g_s\,,
  \nonumber\\
  G_\mu^{0,a}&=&\sqrt{Z_3}G_\mu^a\,,
  \nonumber\\
  c^{0,a}&=&\sqrt{\tilde{Z}_3}c^a\,,
  \nonumber\\
  \Gamma_{ccg}^{0} &=& \tilde{Z}_1 \Gamma_{ccg}\,,
  \label{eq::renconst}
\end{eqnarray}
where $\mu$ is the renormalization scale and $D=4-2\epsilon$ is the
number of space-time dimensions.  From Eq.~(\ref{eq::renconst}) one
obtains a relation that determines the renormalization constant of the
strong coupling:
\begin{eqnarray}
  Z_g &=& \frac{\tilde{Z}_1}{\tilde Z_3\sqrt{Z_3}}
  \,.
  \label{eq::Zg}
\end{eqnarray}
We have explicitly computed the two-loop expression for $Z_g$
by evaluating $Z_3$, $\tilde{Z}_3$ and
$\tilde{Z}_1$ to the corresponding order in the MSSM. The result will be
given in Section~\ref{sec::renconst}.

The finite decoupling coefficient is obtained via
\begin{eqnarray}
  \zeta_g = \frac{Z_g}{Z_g^\prime} \zeta_g^0
  \,,
  \label{eq::zetagren}
\end{eqnarray}
where $Z_g^\prime$ corresponds to the renormalization constant in the
effective theory.

In Ref.~\cite{Chetyrkin:1997un,Steinhauser:2002rq} the framework
described above has been applied to the computation of the three-loop
decoupling coefficients in QCD.  The same method can be applied to the
MSSM with the complication that more than one mass scale appears. In
particular, one has to deal with a massive gluino and different masses
for the two squarks of the same flavour.  Furthermore, the different
hierarchy structures among these masses lead to different expressions
for the decoupling coefficients.  The starting point of all scenarios
defined in the Introduction is the energy regime where all particles of
the theory are active (i.e., they contribute to the $\beta$ function of
the strong coupling); the corresponding coupling is denoted by
$\asfull$.  The end point is given by the regime where only
the five lightest quark flavours are active, with the corresponding
coupling denoted by $\alpha_s^{(5)}$.  In case~(D) a one-step
decoupling from $\asfull$ to $\alpha_s^{(5)}$ is performed. However,
in order to have more flexibility, in scenarios~(A), (B) and (C) a
two-step decoupling is defined
where the corresponding strong couplings and decoupling coefficients are
defined through
\begin{align}
  &\mbox{(A)}&
  \asA               &\,\, =\,\, \left(\zeta_g^{\rm A1}\right)^2 \, \asfull \,,
  &\alpha_s^{(5)} &\,\,=\,\, \left(\zeta_g^{\rm A2}\right)^2 \, \asA    \,,&&&&
  \nonumber\\
  &\mbox{(B)}&
  \asB                &\,\,=\,\, \left(\zeta_g^{\rm B1}\right)^2 \, \asfull \,,
  &\alpha_s^{(5)} &\,\,=\,\, \left(\zeta_g^{\rm B2}\right)^2 \, \asB    \,,&&&&
  \nonumber\\
  &\mbox{(C)}&
  \asC                &\,\,=\,\, \left(\zeta_g^{\rm C1}\right)^2 \, \asfull
  \,,\qquad
  &\alpha_s^{(5)} &\,\,=\,\, \left(\zeta_g^{\rm C2}\right)^2 \, \asC    \,,&&&&
  \nonumber\\
  &\mbox{(D)}&
  \alpha_s^{(5)} &\,\,=\,\, \left(\zeta_g^{\rm D}\right)^2  \, \asfull \,.&&&&
  \label{eq::ABCD}
\end{align}
The dependence on the renormalization scale $\mu$ and --- for
the decoupling coefficients --- on the masses is suppressed.
The main goal of this paper is the computation of 
the $\zeta_g^X$'s.

Due to the isospin constraint between the top and bottom squark masses
(see, e.g., Ref.\,\cite{Heinemeyer:2004xw}), scenario~(A) is
inconsistent. However, from the theoretical point of view it constitutes
an appealing mass pattern. Furthermore, combined with scenario~(C), it
can be used to construct more realistic settings as will be described
briefly at the end of Section~\ref{sec::decconst}.  As our default
scenario in the numerical discussion we adopt scenario~(C).

Dimensional regularization (DREG) is the method which is most widely
used in order to regularize the divergences of loop integrals.  However,
it is well-known that DREG violates supersymmetry. A possible compromise
is provided by dimensional reduction (DRED)~\cite{DRED} which enforces
the equality of fermionic and bosonic degrees of freedom while preserving
all the calculational advantages of DREG.  All the results of this paper
are thus evaluated in the $\overline{\rm DR}$ scheme, i.e., we used DRED
accompanied by minimal subtraction.  

The final results are expressed in terms of the strong coupling constant
$\alpha_s$ in the $\overline{\rm DR}$ scheme and the on-shell mass of
quarks, squarks and gluinos.  The experimental input, $\alpha_s(M_Z)$,
is usually given in the $\overline{\rm MS}$ scheme, however, such that
we need to convert it to the $\overline{\rm DR}$ scheme. We evaluated
the corresponding relation through two loops in the MSSM, finding
\begin{eqnarray}
  \asDRbar &=& \asMSbar\left[1+\frac{\asMSbar}{\pi} \frac{C_A}{12}
  +\left(\frac{\asMSbar}{\pi}\right)^2\left(
  \frac{11}{72}C_A^2 - \frac{1}{16} C_A^2 n_{\tilde{g}} - \frac{1}{8}
  C_F T n_f \right)
  + {\cal O}(\alpha_s^3) \right]\!,\quad
  \label{eq::asMS2DR}
\end{eqnarray}
where $n_f$ is the number of quark flavours and $n_{\tilde{g}}=1$ or $0$
distinguishes whether the gluino is active or not. Only the QCD-part of this
equation will be used in what follows. It is obtained by setting 
$n_{\tilde{g}}= 0$ and agrees with Ref.~\cite{Bern:2002zk}.
With $n_f=5$, the input $\asMSbar(M_Z)=0.1187\pm0.002$~\cite{PDG} gives
$\asDRbar(M_Z)=0.1198\pm0.002$ when using the one-loop approximation of
Eq.~(\ref{eq::asMS2DR}), and $\asDRbar(M_Z)=0.1200\pm0.002$ with
two-loop accuracy.

Note that, in what follows, we will often use the short hand notation
\begin{equation}
\begin{split}
\alpha_s\equiv \alpha_s^{\overline{\rm DR}}(\mu)\,.
\end{split}
\end{equation}

\section{\label{sec::renconst}Renormalization constants and $\beta$ function}

\begin{figure}[t]
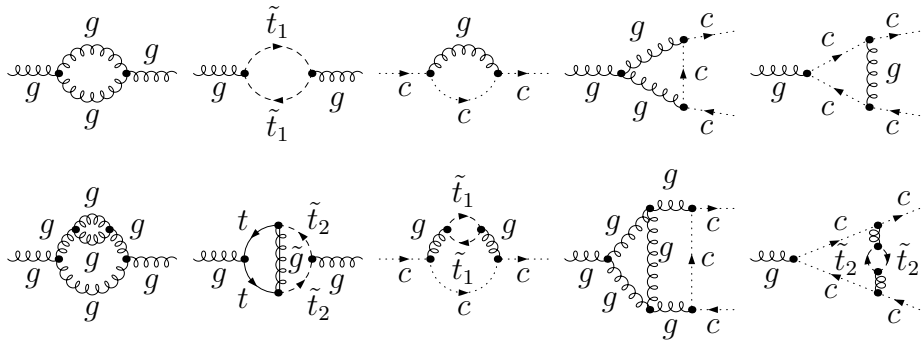

  \begin{center}
    \begin{tabular}{c}
\unitlength=1.bp%
\begin{feynartspicture}(350,140)(5,2)
\input  figs/diagrams
\end{feynartspicture}  
    \end{tabular}
    \parbox{14.cm}{
      \caption[]{\label{fig::diagrams}\sloppy Sample diagrams
        contributing to $Z_3$, $\tilde{Z}_3$ and $\tilde{Z}_1$ with
        gluons ($g$), ghosts ($c$), top quarks ($t$), top squarks
        ($\tilde{t}$) and gluinos ($\tilde{g}$).  The diagrams
        contributing to the decoupling coefficients must contain at
        least one heavy particle.  }}
  \end{center}
\end{figure}

In this section we present the results for the $\overline{\rm DR}$
renormalization constants. Since the $\overline{\rm DR}$ scheme is
mass-independent, the calculation can be reduced to massless integrals
depending on only one external momentum.  More precisely, for $Z_3$,
$\tilde{Z}_3$ and $\tilde{Z}_1$ one has to consider the gluon
propagator, the ghost propagator, and the gluon-ghost vertex; in the
latter case, the external momentum is chosen to flow through the ghost
line.  Sample diagrams are shown in Fig.~\ref{fig::diagrams}.  An
important check is provided by the QCD result which we re-derive for
general gauge parameter $\xi$, defined through the gluon propagator
\begin{eqnarray}
  D_g^{\mu\nu}(q) &=& -i\,
  \frac{g^{\mu\nu}- \xi \frac{q^\mu q^\nu}{q^2}}
  {q^2+ i\varepsilon}
  \,.
\end{eqnarray}
Up to two-loop order our results read~\cite{Capper:1979ns}

\begin{eqnarray}
  Z_3 &=&
  1 + \frac{\alpha_s}{\pi}\Bigg\{ \frac{1}{\epsilon} \bigg[
    C_A \left(\frac{5}{12} -\frac{n_{\tilde{g}}}{6} + \frac{\xi}{8}\right)
    -\frac{T}{6} (2 n_f+n_s)
    \bigg]\Bigg\}
    \nonumber\\&&
  + \left(\frac{\alpha_s}{\pi}\right)^2  \! \Bigg\{
  \frac{1}{\epsilon^2} \Bigg[
    C_A^2\left(-\frac{25}{192}
    + \frac{n_{\tilde{g}}}{96} (5-2\xi)
    +\frac{5\xi}{384}+\frac{\xi^2}{64}
\right)
    + \frac{C_A T}{96} (2 n_f+n_s) (5-2\xi)
    \Bigg]
  \nonumber\\&&\mbox{}\quad
  +\frac{1}{\epsilon} \Bigg[
    C_A^2\left(\frac{25}{128}
    - \frac{9}{64} n_{\tilde{g}}
    +\frac{15\xi}{256}
    -\frac{\xi^2}{128}\right)
    - \frac{C_F T}{8} (n_f+2 n_s - n_s n_{\tilde{g}})
    \nonumber\\&&\mbox{}\qquad
    - \frac{C_A T}{64} (10 n_f -n_s - 8 n_s n_{\tilde{g}})
    \Bigg]
   \Bigg\}\,,\nonumber
\\
  \tilde{Z}_3 &=&
  1 + \frac{\alpha_s}{\pi}\Bigg[\frac{1}{\epsilon} C_A
  \left(\frac{1}{8}+\frac{\xi}{16}\right)
    \Bigg]\nonumber\\
 & &   + \left(\frac{\alpha_s}{\pi} \right)^2 \Bigg\{
    \frac{1}{\epsilon^2} \left[
    C_A^2\left(-\frac{1}{16} + \frac{1}{64} n_{\tilde{g}}
  -\frac{3\xi}{256}+\frac{3\xi^2}{512}\right)
   +C_A T (2 n_f+n_s)\frac{1}{64}\right] \nonumber\\
   & &\quad + \frac{1}{\epsilon} \left[
   C_A^2\left(\frac{43}{768} -\frac{5}{384} n_{\tilde{g}}
   - \frac{\xi}{512}\right)
   - C_A T (10 n_f +11 n_s)\frac{1}{384}\right]
   \Bigg\}
  \,, \nonumber\\
  \tilde{Z}_1 &=&
  1 + \frac{\alpha_s}{\pi}\Bigg[\frac{1}{\epsilon}C_A
           \left(-\frac{1}{8}+\frac{\xi}{8}\right)\Bigg]
\nonumber\\ & &
    + \left(\frac{\alpha_s}{\pi}\right)^2  \Bigg[
      \frac{1}{\epsilon^2} C_A^2
       \left(\frac{5}{128}-\frac{7\xi}{128}
    + \frac{\xi^2}{64}\right)
+  \frac{1}{\epsilon} C_A^2
     \left(-\frac{3}{128}+\frac{7\xi}{256}-\frac{\xi^2}{256}\right)
 \Bigg]
  \,,
  \label{eq::renconst_res}
\end{eqnarray}
where $n_f$ and $n_s$ are the numbers of active quark and squark
flavours, respectively, and $n_{\tilde{g}}=0$ or $1$ distinguishes
whether the gluino is active or not. For example, in the full MSSM it is
$n_f=n_s=6$ and $n_{\tilde g}=1$.  The terms involving $n_s n_{\tilde{g}}$ stem
from diagrams containing a quark-squark-gluino vertex.\footnote{ Note
that in this paper $\tilde q_L$, $\tilde q_R$ are either both active or
both integrated out. Also, we always assume that, if $\tilde q$ is
active, so is $q$ ($q\in \{u,d,c,s,t,b\}$).} The precise meaning of
$\alpha_s$ depends on the actual values of these constants. E.g., for
$n_f=5$, $n_s=n_{\tilde{g}}=0$ it is $\alpha_s=\alpha_s^{(5)}$.  For a
gauge group SU($N_c$), the colour factors in
Eq.~(\ref{eq::renconst_res}) are given by
\begin{equation}
\begin{split} 
C_F=\frac{N_c^2-1}{2N_c}\,,\qquad
C_A=N_c\,,\qquad T=1/2\,.
\end{split}
\end{equation}

The combination of $Z_3$, $\tilde{Z}_3$ and $\tilde{Z}_1$ according to 
Eq.~(\ref{eq::Zg}) leads to the renormalization constant for the
strong coupling
\begin{eqnarray}
  (Z_g)^2 &=& 1 + \frac{\alpha_s}{\pi}\cdot\frac{1}{\epsilon}
  \left[C_A\left( -\frac{11}{12} 
    + \frac{1}{6} n_{\tilde{g}}
    \right)
    + \frac{T}{6} (2 n_f+n_s)\right]\ 
 \nonumber\\\mbox{}
  & & +\left(\frac{\alpha_s}{\pi}\right)^2\Bigg\{
   \frac{1}{\epsilon^2} \left[
      C_A^2\left( -\frac{121}{144} - \frac{5}{18}n_{\tilde g}\right)
  + \frac{T^2}{36} (2 n_f+n_s)^2 \right] 
   \nonumber\\\mbox{}&&\quad
   +\frac{1}{\epsilon} \Bigg[
   C_A^2 \left(-\frac{17}{48} + \frac{1}{6} n_{\tilde{g}}\right) 
   + \frac{C_F T}{8} (n_f+2 n_s -n_s
  n_{\tilde{g}})
   \nonumber\\\mbox{}&&\qquad
   + \frac{C_A T}{24} \left(5 n_f + n_s - 3 n_s n_{\tilde{g}}\right)
  \Bigg]\Bigg\} 
\end{eqnarray}
which is needed for the renormalization of the decoupling coefficient
(cf. Eq.~(\ref{eq::zetagren})).

From $Z_g$ one obtains the $\beta$ function defined through
\begin{eqnarray}
  \mu^2 \frac{{\rm d}}{{\rm d} \mu^2}\frac{\alpha_s}{\pi}
  \,\,=\,\,
  \beta(\alpha_s) &=&
  - \left(\frac{\alpha_s}{\pi}\right)^2
  \sum_{i\ge0} \beta_i \left(\frac{\alpha_s}{\pi}\right)^i
  \,,
\end{eqnarray}
with the result
\begin{eqnarray}
  \beta_0 &=& \frac{1}{4}\left[
    C_A\left(\frac{11}{3}-\frac{2}{3} n_{\tilde{g}}\right)
 -\frac{2}{3} T (n_s+ 2 n_f)\right]
  \,,
\\
  \beta_1 &=& \frac{1}{16}\left[
    C_A^2\left(\frac{34}{3} -\frac{16}{3} n_{\tilde{g}}\right) 
  - \frac{4}{3}C_A T ( 5 n_f + n_s  - 3 n_s n_{\tilde{g}} )
  - 4C_F T(n_f +2 n_s -n_s n_{\tilde{g}}) \right]
  \nonumber  \,.
\end{eqnarray}
Note that due to the terms $\propto n_s n_{\tilde{g}}$ this result can also be
used for scenarios where the gluino is decoupled but (some) squarks are
not. However, for the three-loop coefficient $\beta_2$ (see below) such
terms are not separately available.  Thus in the following we identify $n_s
n_{\tilde{g}} \to n_s$.

The results for $\beta_0$ and $\beta_1$ are renormalization scheme
independent. The pure QCD part agrees with the well-known $\overline{\rm
MS}$ result which can be found in Ref.~\cite{Muta}, for example.  The
contribution from the SUSY particles agrees with
Ref.~\cite{Jack:1996vg}.

The two-loop decoupling relations presented in the next section have to
be used in conjunction with the three-loop coefficient of the $\beta$
function. Subtracting the pure QCD result in the $\overline{\rm DR}$
scheme~\cite{Bern:2002zk} from the fully supersymmetric
expression~\cite{Jack:1996vg}, we can distinguish the contributions from
squarks and gluinos and multiply them by the appropriate labels $n_s$
and $n_{\tilde g}$:
\begin{eqnarray}
  \beta_2 &=& \frac{1}{64}\bigg[ C_A^3\left( \frac{3115}{54} -
    \frac{1981}{54} n_{\tilde{g}} \right) 
    \!+\! C_A^2 T \left(\frac{899}{27} n_s -
    \frac{1439}{27} n_f \right)
     \!+\! C_A T^2 \left(\!- \frac{50}{27} n_s^2 +
    \frac{158}{27} n_f^2 \right) 
    \nonumber\\&& 
   - C_A C_F T \left(\frac{209}{9} n_s +
    \frac{259}{9} n_f \right)
    + C_F^2 T \left(14 n_s+2 n_f\right) 
    + C_F T^2 \left(\frac{148}{9} n_s^2 +
    \frac{68}{9} n_f^2 \right)
    \bigg]
  \,.
  \nonumber\\
  \label{eq::beta2}
\end{eqnarray}
Note that since Ref.~\cite{Jack:1996vg} assumed $n_f=n_s$, it is not
possible to distinguish the term proportional to $n_s n_f$ from the
$n_s^2$ term. For this reason, Eq.\,(\ref{eq::beta2}) can only be
used for $n_f=n_s$ or $n_s=0$; unfortunately, this prevents us from
discussing scenario (A) in the numerical applications. As expected, the
pure QCD result in $\overline{\rm DR}$, obtained by setting
$n_s=n_{\tilde{g}}=0$ in Eq.\,(\ref{eq::beta2}), is different from the
expression in the $\overline{\rm MS}$ scheme; only the coefficients of
the colour structures $C_AT^2n_f^2$ and $C_F^2Tn_f$ are the same in both
schemes.

Apart from the $\overline{\rm DR}$ renormalization constants in
Eq.~(\ref{eq::renconst_res}), we also need the one-loop renormalization
constants for the top quark and squark mass.  The explicit on-shell
expressions are given in Eqs.~(B.5) and~(B.6) of
Ref.~\cite{Harlander:2004tp}.  The on-shell counterterm for the gluino
mass reads (see, e.g., Ref.~\cite{Pierce:1996zz})
\begin{eqnarray}
  \frac{m_{\tilde{g}}^B}{m_{\tilde{g}}} &=& 1- \frac{\alpha_s}{ \pi}\Bigg\{
  \frac{1}{4 \epsilon}(3 C_A - 2 T n_s) 
  +\frac{1}{2} C_A \left(  \frac{5}{2} + \frac{3}{2}L_{\tilde{g}}\right)
  \nonumber\\
  & & + T \sum_{q}\sum_{i=1}^2\frac{1}{4}\left[
    \frac{m_q^2}{m_{\tilde{g}}^2} \left(1+L_q\right)
    -\!\frac{m_{\tilde{q}_i}^2}{m_{\tilde{g}}^2}
    \left(1+L_{\tilde{q}_i}\right)
    +\frac{m_{\tilde{q}_i}^2 - m_{\tilde{g}}^2-
      m_q^2}{m_{\tilde{g}}^2} 
    B_0^{\rm fin}(m_{\tilde{g}}^2, m_q,m_{\tilde{q}_i})\right]
  \nonumber\\
  & &  - T\sum_q \frac{1}{2}\frac{m_q}{m_{\tilde{g}}}\sin 2\theta_q
  \left[B_0^{\rm fin}
    (m_{\tilde{g}}^2,m_q,m_{\tilde{q}_1})
    -B_0^{\rm fin}(m_{\tilde{g}}^2,m_q,m_{\tilde{q}_2})\right]
  \Bigg\}
  \,,
  \label{eq::mgluino}
\end{eqnarray}
where $B_0^{\rm fin}$ is defined in Eq.~(B.8) of
Ref.~\cite{Harlander:2004tp}.  Note that the ${\cal O}(\epsilon)$ part
of the mass counterterms is not needed since the one-loop result depends
only logarithmically on the masses.  In Eq.~(\ref{eq::mgluino}), we have
used the short-hand notation
\begin{equation}
\begin{split}
L_x &= \ln\frac{\mu^2}{m_x^2}\,,\qquad x\in \{q,\,t,\,\tilde g,\,\tilde
q_i,\,\ldots\}\,.
\label{eq::logdef1}
\end{split}
\end{equation}
In addition to that, the abbreviations
\begin{equation}
\begin{split}
L_{\tilde M} &= \ln\frac{\mu^2}{\tilde M^2}\,,\qquad
L_{\tilde m} = \ln\frac{\mu^2}{\tilde m^2}\,,\qquad
L_{t\tilde m} = \ln\frac{m_t^2}{\tilde m^2}\,,
\label{eq::logdef2}
\end{split}
\end{equation}
will be used in what follows.

\section{\label{sec::decconst}Decoupling coefficients}

As already pointed out in Section~\ref{sec::tech}, the calculation of
the decoupling coefficients can be reduced to vacuum
integrals~\cite{Chetyrkin:1997sg,Chetyrkin:1997un,Steinhauser:2002rq}.
At two-loop level, such integrals can be evaluated fully
analytically~\cite{Davydychev:1992mt} for arbitrary masses. For the
purpose of this paper, however, it is sufficient to solve them for
either a single or for two largely separated mass scales. In the first
case, the analytical result is rather short in general, while in the
second case we can perform an asymptotic expansion in the ratio of the
two mass scales and truncate the series after the first few terms which
leads again to a fairly compact expression.  Furthermore, in order to
shorten the expressions even more, we present the results for the SU(3)
gauge group.

The Feynman diagrams are generated using {\tt
  QGRAF}~\cite{Nogueira:1991ex}, and subsequently translated by {\tt
  Q2E/EXP}~\cite{Seidensticker:1999bb,Harlander:1997zb} to the notation
  of {\tt MATAD}~\cite{Steinhauser:2000ry}, which evaluates the vacuum
  integrals. If the integrals contain more than one mass scale, {\tt
  EXP} performs an asymptotic expansion on the diagrams before passing
  them on to {\tt MATAD}.

Before discussing the different scenarios, we present the pure QCD
result for $\zeta_g$ in the $\overline{\rm DR}$ scheme, where the top
quark is considered as heavy and the remaining $(n_f-1)=5$ quarks are
massless.  Up to two-loop order, the decoupling coefficient relating
$\alpha_s^{(n_f)}$ and $\alpha_s^{(n_f-1)}$ via
\begin{eqnarray}
  \alpha_s^{(n_f-1)} &=& (\zeta_g^{\rm QCD})^2 \alpha_s^{(n_f)}
  \label{eq::aseff}
\end{eqnarray}
reads
\begin{eqnarray}
  (\zeta_g^{\rm QCD})^2 &=& 1 - \frac{\alpha_s^{(n_f)}}{\pi} \frac{1}{6} L_t +
  \left(\frac{\alpha_s^{(n_f)}}{\pi}\right)^2  \left(-\frac{5}{24}
  -\frac{19}{24}  L_t + \frac{1}{36}  L_t^2\right) 
  +  {\cal O}\left(\alpha_s^3\right)
  \,,
  \label{eq::zetagQCD}
\end{eqnarray}
with $L_t=\ln (\mu^2/m_t^2)$.  Note that in the $\overline{\rm MS}$
scheme the constant term of order $\alpha_s^2$ is $-7/24$ instead of
$-5/24$ as in
Eq.~(\ref{eq::zetagQCD})~\cite{Bernreuther:1981sg,Larin:1994va,Chetyrkin:1997sg,Chetyrkin:1997un}.
The transformation of Eq.~(\ref{eq::aseff}) from the $\overline{\rm DR}$
to the $\overline{\rm MS}$ scheme confirms the fermionic two-loop
coefficient of Eq.~(\ref{eq::asMS2DR}).

Within the MSSM, the one-loop result for the decoupling coefficient
relating $\alpha_s^{(5)}$ to the strong coupling where also squarks and
gluinos are active can be written in the generic form (see, e.g.,
Ref.~\cite{Harlander:2004tp})
\begin{eqnarray} 
  (\zeta_g^{\rm SUSY})^2 &=& 1 + \frac{\alpha_s^{(\rm SUSY)}}{\pi}
  \Bigg[-
  \frac{1}{6} L_t -  \frac{1}{12} \sum_{s=1}^{n_s} L_{\tilde{q}_s} -
  \frac{n_{\tilde{g}}}{2}
  L_{\tilde{g}}
  - \epsilon \Bigg(
  \frac{1}{12} \left(\zeta(2)+ L_t^2\right)
  \nonumber\\
  &&\mbox{}  
  + \frac{1}{24}\sum_{s=1}^{n_s}\left(\zeta(2)+   L_{\tilde{q}_s}^2\right)
  + \frac{n_{\tilde{g}}}{4} \left(\zeta(2)+ L_{\tilde{g}}^2\right) \Bigg)
  \Bigg]
  + {\cal O}\left(\alpha_s^2\right)
  \,,
  \label{eq::zetag1l}
\end{eqnarray}
with the notation as defined in Eqs.~(\ref{eq::logdef1})
and~(\ref{eq::logdef2}), and $\zeta(2)=\pi^2/6=1.64493\ldots$.  For the
sake of consistency, also the term of order $\epsilon$ is given; it is
needed in intermediate steps of the calculation.  The sum runs over all
active squark flavours. We introduced a generic label ``SUSY'' which has
to be adapted to the scenarios (A)--(D) discussed above.  Note that due
to the different mass scales involved in Eq.~(\ref{eq::zetag1l}),
$\alpha_s$ exhibits a jump even at one-loop order if all heavy particles
are integrated out at the same scale.

Let us now discuss the various cases defined in the Introduction and in
Eq.~(\ref{eq::ABCD}). The common large mass will be denoted by
$\tilde{M}$ and the common smaller mass by $\tilde{m}$.  In some cases,
the SUSY masses will be taken to be (much) larger than the top quark
mass. However, it turns out that the expansion in the ratio converges
very fast --- even for equal masses.  Since in Section~\ref{sec::num}
the inverse decoupling coefficients are needed, we will give explicit
results for them. The analytic expressions for $\zeta_g$ itself can be
obtained easily by recursive substitution of the one-loop relations.

\subsection{Scenario~(A)}
\[
\begin{split}
m_{\tilde{u}} = \ldots = m_{\tilde{b}} = \tilde{M}\quad &\gg\quad
\tilde{m} = m_{\tilde{t}} = m_{\tilde{g}}\,,\ m_t\nonumber\\ \asA =
(\zeta_g^{\rm A1})^2 \, \asfull \,,&\qquad \alpha_s^{(5)} =
(\zeta_g^{\rm A2})^2 \,
\asA
\end{split}
\]
$\zeta_g^{\rm A1}$ is a function of $\tilde{M}$ only. In fact, following
the method of
Ref.~\cite{Chetyrkin:1997sg,Chetyrkin:1997un,Steinhauser:2002rq}, the
masses of the gluino, the top squarks and all quarks have to be set to
zero in the corresponding Feynman diagrams,\footnote{Note that this is
no restriction on the resulting effective theory: once the heavy
particles are decoupled, the remaining fields can have arbitrary masses
$\ll \tilde M$.} cf.\ Fig.\,\ref{fig::diagrams}.  The result is
\begin{eqnarray}
  \frac{1}{(\zeta_g^{\rm A1})^2} &=& 1 + \frac{\asA}{\pi} \frac{n_s}{12} 
L_{\tilde{M}} + \left(\frac{\asA}{\pi}\right)^2  \left(-\frac{13}{48} n_s  
-\frac{n_s}{12} L_{\tilde{M}}+ \left(\frac{n_s}{12}
L_{\tilde{M}}\right)^2\right)
  \,,
\end{eqnarray}
with $n_s=5$.

The Feynman diagrams leading to $\zeta_g^{\rm A2}$ do not contain 
the five squark flavours already integrated out in step A1.
However, one has the top squarks, gluino and top quark as massive. 
We consider two limits:\\[1em]
\underline{(A2a) $\tilde m = m_t$:}
\begin{align}
\frac{1}{(\zeta_g^{\rm A2a})^2} &= 1 + \frac{\alpha_s^{(5)}}{\pi} \frac{3}{4}
 L_{\tilde{m}}
 \nonumber\\&
 + \left(\frac{\alpha_s^{(5)}}{\pi} \right)^2
 \left( 
 \frac{1909}{864}
 + \frac{11\sqrt{3}}{108}\pi  + \frac{17}{12} S_2
 + \frac{89}{24} L_{\tilde m} + \left(\frac{3}{4} L_{\tilde m}\right)^2
 \right) 
  \,,
\end{align}
\mbox{}\\
\underline{(A2b) $\tilde m\gg m_t$:}
\renewcommand{\lM}{L_{\tilde m}}
\renewcommand{\lt}{L_{t}}
\renewcommand{\ltM}{L_{t\tilde m}}
\begin{align}
\frac{1}{(\zeta_g^{\rm A2b})^2} &= 1 
+ \frac{\alpha_s^{(5)}}{ \pi}\left(
\frac{1}{6} L_t +\frac{7}{12}  L_{\tilde{m}} 
\right)
\nonumber\\
 & + \left(\frac{\alpha_s^{(5)}}{\pi}\right)^2
 \bigg[
        \frac{265}{96} 
        + \frac{19}{24}\lt 
        + \frac{35}{12}\lM 
        + \left(
        \frac{1}{6} L_t 
        + \frac{7}{12}  L_{\tilde{m}} 
        \right)^2
        \nonumber\\&\quad
        + \left(\frac{m_t}{\tilde m}\right)^2 \left(  
        - \frac{5}{48} 
        - \frac{3}{8}\ltM 
        \right)
        +  \frac{7\pi}{36} \left(\frac{m_t}{\tilde m}\right)^3
        \nonumber\\&\quad
        + \left(\frac{m_t}{\tilde m}\right)^4 \left(  
        - \frac{881}{7200} 
        + \frac{1}{80}\ltM 
        \right)
        - \frac{7\pi}{288} \left(\frac{m_t}{\tilde m}\right)^5
        + \ldots
        \bigg]\,,
  \label{eq::izetagA2b}
\end{align}
where $S_2 = \frac{4}{9 \sqrt{3}}{\rm Cl}_2(\frac{\pi}{3}) \simeq 0.260434$.
It is interesting to note that
$1/(\zeta_g^{\rm A2b})^2$ as given in Eq.~(\ref{eq::izetagA2b}) 
evaluated for $\tilde{m}=m_t$ 
approximates $1/(\zeta_g^{\rm A2a})^2$ to an accuracy
better than 1\%.

\subsection{Scenario~(B)}
\[
\begin{split}
m_{\tilde{u}} = \ldots = m_{\tilde{t}} = \tilde{M} \quad &\gg\quad
 m_{\tilde{g}}\,,\ m_t\\ \asB = (\zeta_g^{\rm B1})^2 \, \asfull \,,&\qquad
 \alpha_s^{(5)} = (\zeta_g^{\rm B2})^2 \, \asB
\end{split}
\]
The analytical expression for $\zeta_g^{\rm B1}$ is identical to the one
of $\zeta_g^{\rm A1}$, only now we have $n_s=6$.  In the case of
$\zeta_g^{\rm B2}$ there is no gluino-top-stop interaction since at this
stage all squarks are already integrated out. As a consequence the
integrations factorize into parts where either the gluino or the top
quark is the only massive particle. One obtains
\renewcommand{\lt}{L_{t}} \renewcommand{\lgl}{L_{\tilde g}}
\begin{align}
  \frac{1}{(\zeta_g^{\rm B2})^2} &= 1 
  + \frac{\alpha_s^{(5)}}{ \pi}
  \left(
  \frac{1}{6} \lt
  + \frac{1}{2} \lgl
  \right)
  \nonumber\\&
  + \left(\frac{\alpha_s^{(5)}}{\pi}\right)^2
  \left[
  \frac{275}{96}
  + \frac{19}{24}\lt 
  + 3\lgl
  + \left(
  \frac{1}{6} \lt
  + \frac{1}{2} \lgl
  \right)^2
  \right]
  \,.
\end{align}

\subsection{Scenario~(C)}
\[
\begin{split}
m_{\tilde{u}} = \ldots = m_{\tilde{t}} = &\ m_{\tilde{g}} = \tilde{M}\quad
  \gg\quad m_t
  \\
  \asC = (\zeta_g^{\rm C1})^2 \, \asfull \,,&\qquad 
  \alpha_s^{(5)} = (\zeta_g^{\rm C2})^2 \, \asC
\end{split}
\]
The result for $\zeta_g^{\rm C1}$ is given by
\renewcommand{\lM}{L_{\tilde M}}
\begin{align}
    \frac{1}{( \zeta_g^{\rm C1})^2} &= 
    1 
    + \frac{\asC}{\pi}
    \left(
    \frac{1}{2}
    + \frac{n_s}{12}
    \right) \lM
    \nonumber\\&
    + \left(
    \frac{\asC}{\pi}\right)^2\bigg[
      \frac{85}{32}
      - \frac{5}{48} n_s
      + \left( 
      3 - \frac{n_s}{12}
      \right) L_{\tilde{M}} 
      + \left(
      \frac{1}{2}
      + \frac{n_s}{12}
      \right)^2 \lM^2
      \bigg]
    \,,
\end{align}
with $n_s=6$. 
$\zeta_g^{\rm C2}$ is identical to the pure QCD result of
Eq.~(\ref{eq::zetagQCD}),
\begin{equation}
\begin{split}
 \zeta_g^{\rm C2} = \zeta_g^{\rm QCD}\,.
\end{split}
\end{equation}

\subsection{Scenario~(D)}
\[
\begin{split}
\tilde{M} = m_{\tilde{u}} = \ldots &= m_{\tilde{t}} = m_{\tilde{g}}
\,, \ m_t\\ \alpha_s^{(5)} &= (\zeta_g^{\rm D})^2 \, \asfull
\end{split}
\]
As in the case (A2) we consider two limits:\\[1em]
\underline{(Da) $\tilde{M}=m_t$:}
\begin{align}
  \frac{1}{\left(\zeta_g^{\rm Da}\right)^2} 
  &= 1 + \frac{\alpha_s^{(5)}}{\pi}
  \left(
  \frac{2}{3}
  + \frac{n_s}{12}
  \right)  L_{\tilde{M}} 
  + \left(\frac{\alpha_s^{(5)}}{\pi}\right)^2
  \bigg[
    \frac{1999}{864} 
    + \frac{11\sqrt{3}}{108} \pi  
    + \frac{17}{12} S_2 
    - \frac{5}{48} n_s
    \nonumber\\&\qquad
    + \left(
    \frac{91}{24} 
    -\frac{n_s}{12}
    \right) L_{\tilde{M}}
    +  \left(
    \frac{2}{3}
    + \frac{n_s}{12}
    \right)^2  L^2_{\tilde{M}}
    \bigg]
  \,,
\end{align}
\mbox{} \\
\underline{(Db) $\tilde{M}\gg m_t$:}
\begin{align}
  \frac{1}{\left(\zeta_g^{\rm Db}\right)^2} 
  &= 1 + \frac{\alpha_s^{(5)}}{\pi}
  \left[
    \frac{1}{6} L_t
    + \left(
    \frac{1}{2} 
    + \frac{n_s}{12}
    \right) L_{\tilde{M}}\right]
    + \left(\frac{\alpha_s^{(5)}}{\pi}\right)^2
    \Bigg\{
      \frac{275}{96}
      - \frac{5}{48} n_s 
      + \frac{19}{24} L_t
      \nonumber\\&\qquad
      + \left(
      3 - \frac{n_s}{12}
      \right) L_{\tilde{M}} 
      + \left[
      \frac{1}{6} L_t
      + \left(
      \frac{1}{2} 
      + \frac{n_s}{12}
      \right ) L_{\tilde{M}}\right]^2
      \nonumber\\&\quad
      + \left(\frac{m_t}{\tilde M}\right)^2 \left(  
      - \frac{5}{48} 
      - \frac{3}{8}\ltM 
      \right)
      +  \frac{7\pi}{36} \left(\frac{m_t}{\tilde M}\right)^3
      \nonumber\\&\quad
      + \left(\frac{m_t}{\tilde M}\right)^4 \left(  
      - \frac{881}{7200} 
      + \frac{1}{80}\ltM 
      \right)
      - \frac{7\pi}{288} \left(\frac{m_t}{\tilde M}\right)^5
      + \ldots
      \Bigg\}
  \,,
  \label{eq::izetagD}
\end{align}
where we have $n_s=6$ in both cases.  We again observe very good
convergence of $1/\left(\zeta_g^{\rm Db}\right)^2$ even for
$\tilde{M}=m_t$; thus, for our numerical analysis in
Section~\ref{sec::num} we only take the result from
Eq.\,(\ref{eq::izetagD}).

At this point two remarks are in order.
First, it is interesting to note that the following relation
holds:
\begin{eqnarray}
  \zeta^{\rm C1} \zeta^{\rm C2} &=& \zeta^{\rm Db} 
  + {\cal O}\left(\frac{m_t^2}{\tilde{M}^2}\right)
  \,,
\end{eqnarray}
which constitutes an important cross check of our calculation.
Second, let us remark that it is possible to 
add the terms of $\zeta_g^{\rm C1}$ involving $n_s$ to
$\zeta_g^{\rm A2a}$ or $\zeta_g^{\rm A2b}$ and thus take into account
additional squarks as being integrated out together with the 
top squark, gluino and top quark.
Of course, the parameter $n_s$ of $\zeta_g^{\rm A1}$ has to be
adjusted accordingly.

\section{\label{sec::num}Numerics}

In this section we study the phenomenological implications of the
two-loop decoupling coefficients.  In particular, we compute the value
of the strong coupling $\alpha_s^{\rm (full)}(M_{\rm SUSY})$ at a high
scale $M_{\rm SUSY}$, as evolved from the experimental input value
$\alpha_s^{(5)}(M_Z)$.  As described in Section~\ref{sec::decoupling},
at higher orders in perturbation theory heavy particles have to be
decoupled properly when crossing the corresponding particle
threshold. However, the precise value of the scale where this is done,
$\mu_{\rm th}$, is not fixed by theory and can be used as a means to
estimate the theoretical uncertainty.  On general grounds one expects that, at
fixed order perturbation theory, the relation between
$\alpha_s^{(5)}(M_Z)$ and $\alpha_s^{(\rm full)}(M_{\rm SUSY})$ is not
very sensitive to the choice of the matching scale as long as $\mu_{\rm
th}$ is of the order of the heavy mass scale.  In the following we study
$\alpha_s^{\rm (full)}(M_{\rm SUSY})$ as a function of $\mu_{\rm th}$
for the first three orders in perturbation theory.

Let us describe the procedure for computing $\alpha_s^{(\rm
  full)}(M_{\rm SUSY})$ from the knowledge of $\alpha_s^{(5)}(M_Z)$ in
detail for scenario (C); the other cases can be treated in complete
analogy.  First, we compute $\alpha_s^{(5)}(\mu_{\rm th}^{\rm C2})$ from
$\alpha_s^{(5)}(M_Z)$ by using the $L$-loop $\beta$ function with
$n_f=5$ and $n_s=n_{\tilde{g}}=0$, and numerically solving the
corresponding renormalization group equation.  In the next step, the
$(L-1)$-loop expression of $\zeta_g^{\rm C2}$ is used to arrive at
$\alpha_s^{(6)}(\mu_{\rm th}^{\rm C2})$.  With the help
of the $L$-loop $\beta$ function (with $n_f=6$, $n_s=0$ and
$n_{\tilde{g}}=0$) we then evaluate $\alpha_s^{(6)}(\mu_{\rm
  th}^{\rm C1})$ and perform the matching to $\alpha_s^{\rm 
  (full)}(\mu_{\rm th}^{\rm C1})$ using $\zeta_g^{\rm C1}$ at $(L-1)$-loop
order. Finally, $\alpha_s^{\rm (full)}(M_{\rm SUSY})$ is obtained
again with the help of the $L$-loop $\beta$ function after setting
$n_f=6$, $n_s=6$ and $n_{\tilde{g}}=1$.  We apply this procedure for
$L=1$, $2$ and $3$ and study the dependence on the two matching scales
$\mu_{\rm th}^{\rm C1}$ and $\mu_{\rm th}^{\rm C2}$.  Of course, in
scenario~(D) there is only one matching scale and the procedure
simplifies accordingly.

Before discussing the numerical effects, let us specify the
input values used in our analysis. In the four scenarios defined in
Section~\ref{sec::decoupling} we need to define a large and a small
mass scale, $\tilde M$ and $\tilde m$, and the top quark mass. We choose
\begin{eqnarray}
  m_t           &=& 174~\mbox{GeV}\,,\nonumber\\
  \tilde M &=& 1000~\mbox{GeV}\,,\nonumber\\
  \tilde m &=& 400~\mbox{GeV}\,.
  \label{eq::numval}
\end{eqnarray}
For the default values of the matching scales we assume $\mu_{\rm
th,1}=\tilde M$ and $\mu_{\rm th,2} = \tilde m$.  Furthermore,
according to the discussion below Eq.~(\ref{eq::asMS2DR}), we use
\begin{eqnarray}
  \alpha_s^{(5)}(M_Z) = 0.120\pm 0.002\,,
  \label{eq::asDRin}
\end{eqnarray}
where the label $\overline{\rm DR}$ has again been omitted.  The GUT
scale will be taken as $\mu_{\rm GUT}=10^{16}$\,GeV.

\begin{table}
  \begin{center}
    \begin{tabular}{c|c}
      $\mu$ & $\alpha_s(\mu)$ \\
      \hline
      800~GeV         & $0.0916 \pm 0.0012$\\
      $\mu_{\rm GUT}$ & $0.0398 \pm 0.00023$
    \end{tabular}
    \parbox{14.cm}{
      \caption[]{\label{tab::alphas}\sloppy
        $\alpha_s(\mu)$ in the $\overline{\rm DR}$ scheme 
        for $\mu=800$~GeV and $\mu=\mu_{\rm GUT}$ assuming scenario (C). 
        The starting point for the prescription described above 
        Eq.~(\ref{eq::numval}) is defined in Eq.~(\ref{eq::asDRin}).
        The error is due to the experimental uncertainty of
        $\alpha_s(M_Z)$.
        }}    
  \end{center}
\end{table}

\begin{figure}[t]
  \begin{center}

    \vspace*{-12em}

    \begin{tabular}{c}
      \epsfig{figure=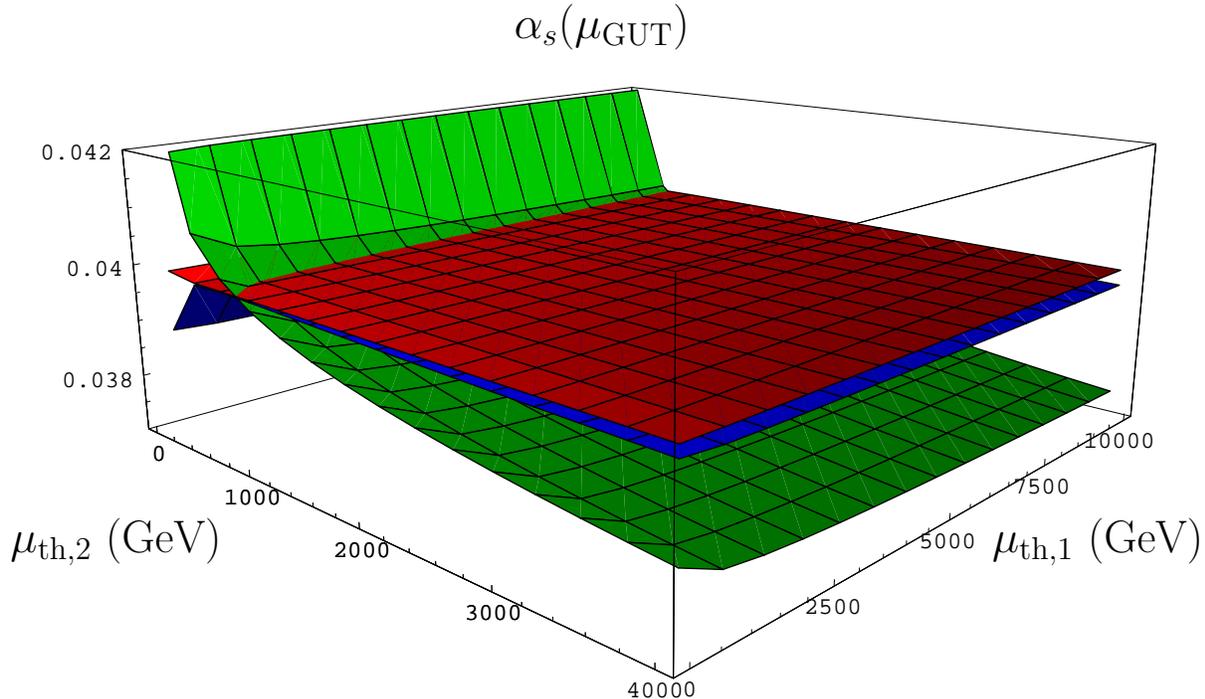,width=36em}
    \vspace*{-10em}
    \end{tabular}

    \vspace*{-28em}

    {\Large $\alpha_s(\mu_{\rm GUT})$}\\[15em]
    \mbox{}\vspace*{2em}
         {\Large $\mu_{\rm th,2}$~(GeV) \hfill $\mu_{\rm th,1}$~(GeV)}

    \vspace*{3em}

    \parbox{14.cm}{
      \caption[]{\label{fig::as3D}\sloppy
$\alpha_s(\mu_{\rm GUT})$ for $\mu_{\rm GUT}=10^{16}$~GeV as a function 
of the two matching scales $\mu_{\rm th,1}$ and $\mu_{\rm th,2}$.
The lower (green), middle (blue) and upper (red) planes correspond to
one-, two- and three-loop running accompanied with the corresponding 
order in the decoupling relation for the scenario (C).
        }}
  \end{center}
\end{figure}

In order to get an impression of the numerical value
for $\alpha_s$ at various scales and the uncertainties
induced by the input $\alpha_s(M_Z)$,
Tab.~\ref{tab::alphas} gives $\alpha_s(\mu)$ for $\mu=800$~GeV and
$\mu=\mu_{\rm GUT}$. These numbers are obtained from Eq.~(\ref{eq::asDRin}) for
scenario~(C) with three-loop running and two-loop matching.  Further on,
the dependence on the matching scales and the loop order is exemplified
in Fig.~\ref{fig::as3D} which shows $\alpha_s(\mu_{\rm GUT})$ for
scenario~(C) as a function of $\mu_{\rm th,1}$ and $\mu_{\rm th,2}$,
each of them varying by a factor of ten around its default value.  One
can clearly see how the result stabilizes as soon as two-loop running
and one-loop matching (middle plane) is included.  The three-loop
corrections stabilize the dependence of $\alpha_s$ on $\mu_{\rm th,1}$
and $\mu_{\rm th,2}$ even more (upper plane), such that the result is
practically independent of the matching scales within a wide range
around their default values. Within the considered range of $\mu_{\rm
th,1}$ and $\mu_{\rm th,2}$, the variation of $\alpha_s(\mu_{\rm GUT})$
is 13.8/1.6/0.3\% at 1-/2-/3-loop order.

\begin{figure}[ht]
  \begin{center}
    \begin{tabular}{cc}
      \epsfig{figure=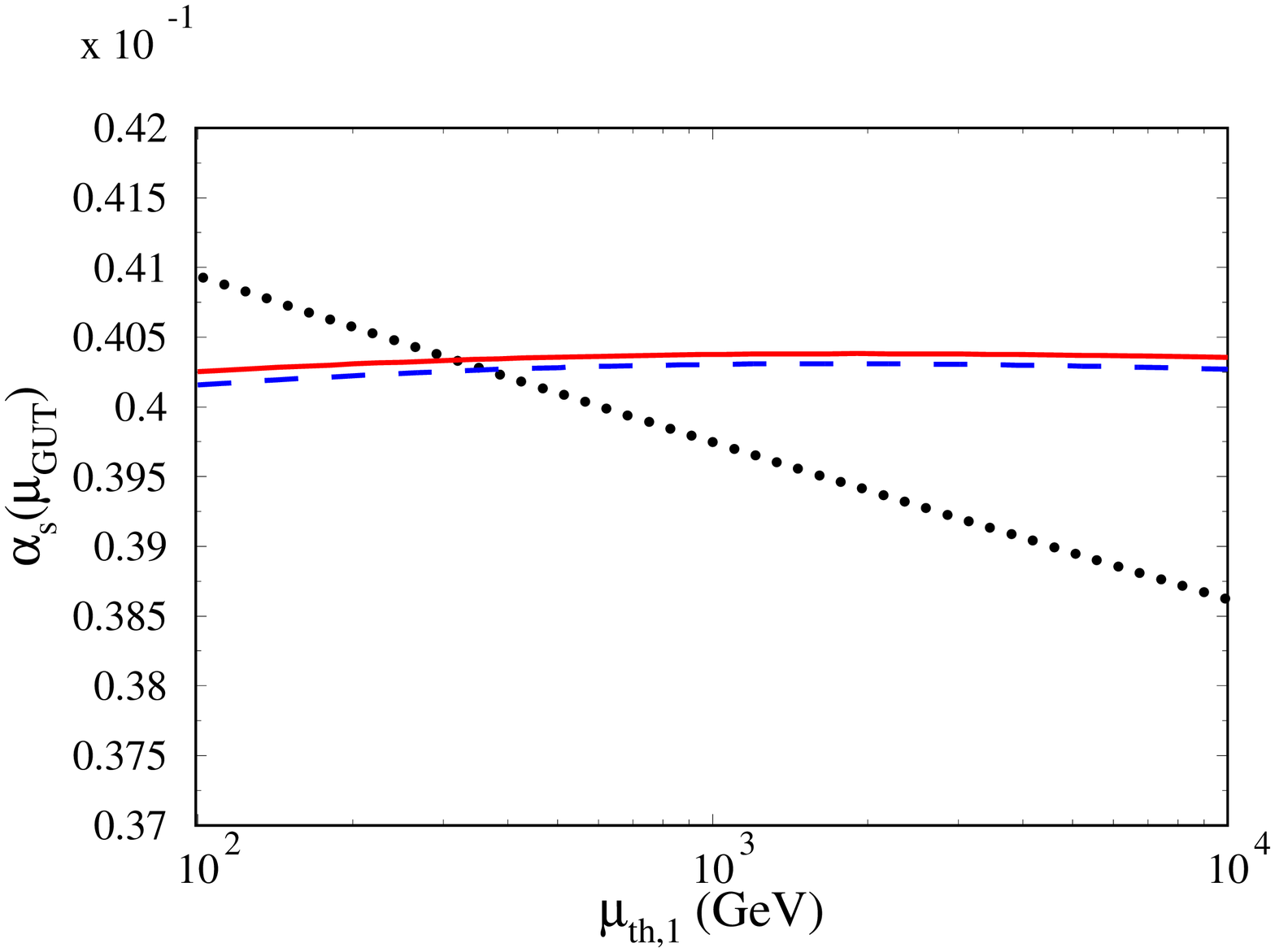,width=18em}
      &
      \epsfig{figure=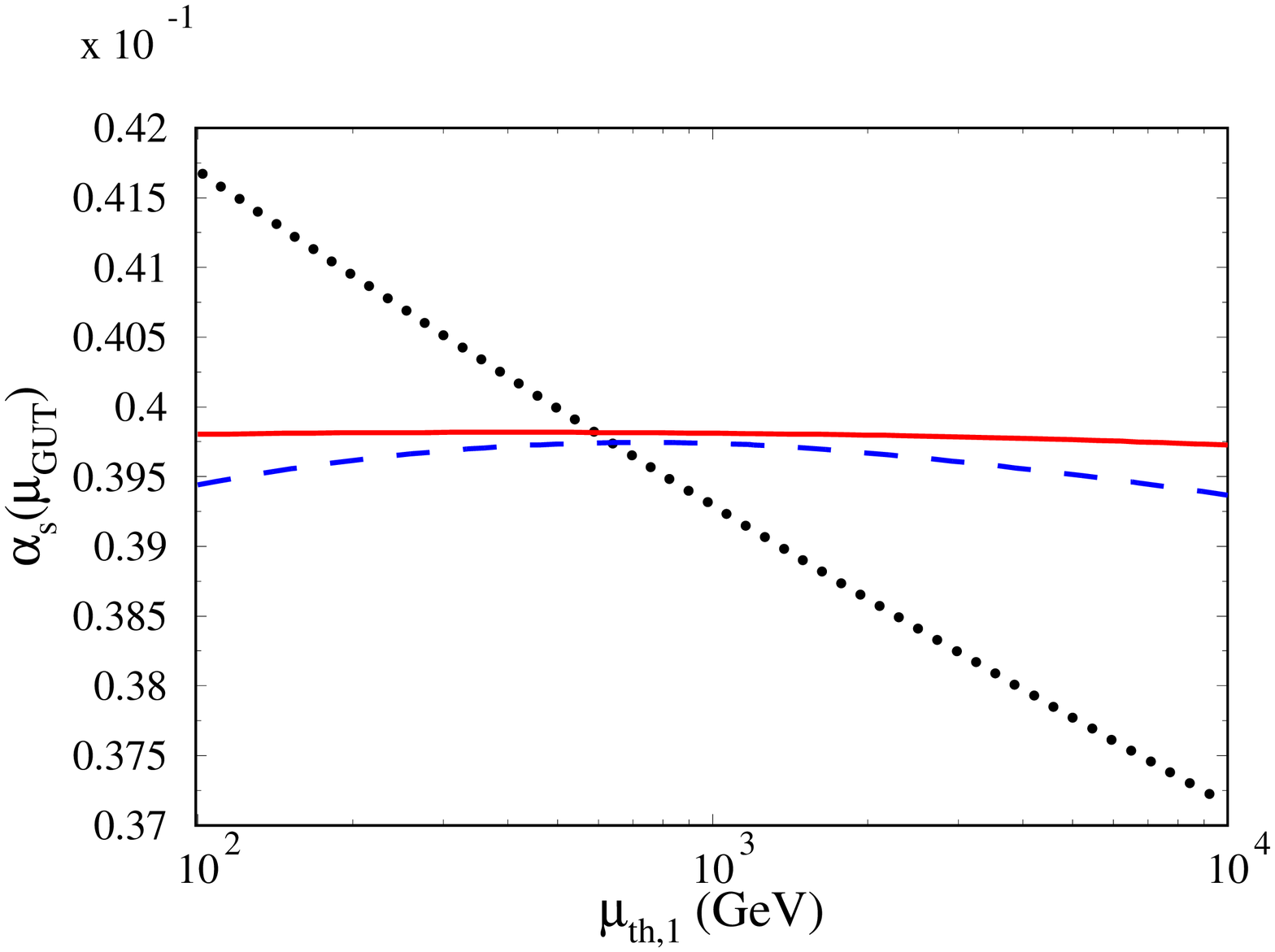,width=18em}
      \\(B) & (C)\\
      \multicolumn{2}{c}{
        \epsfig{figure=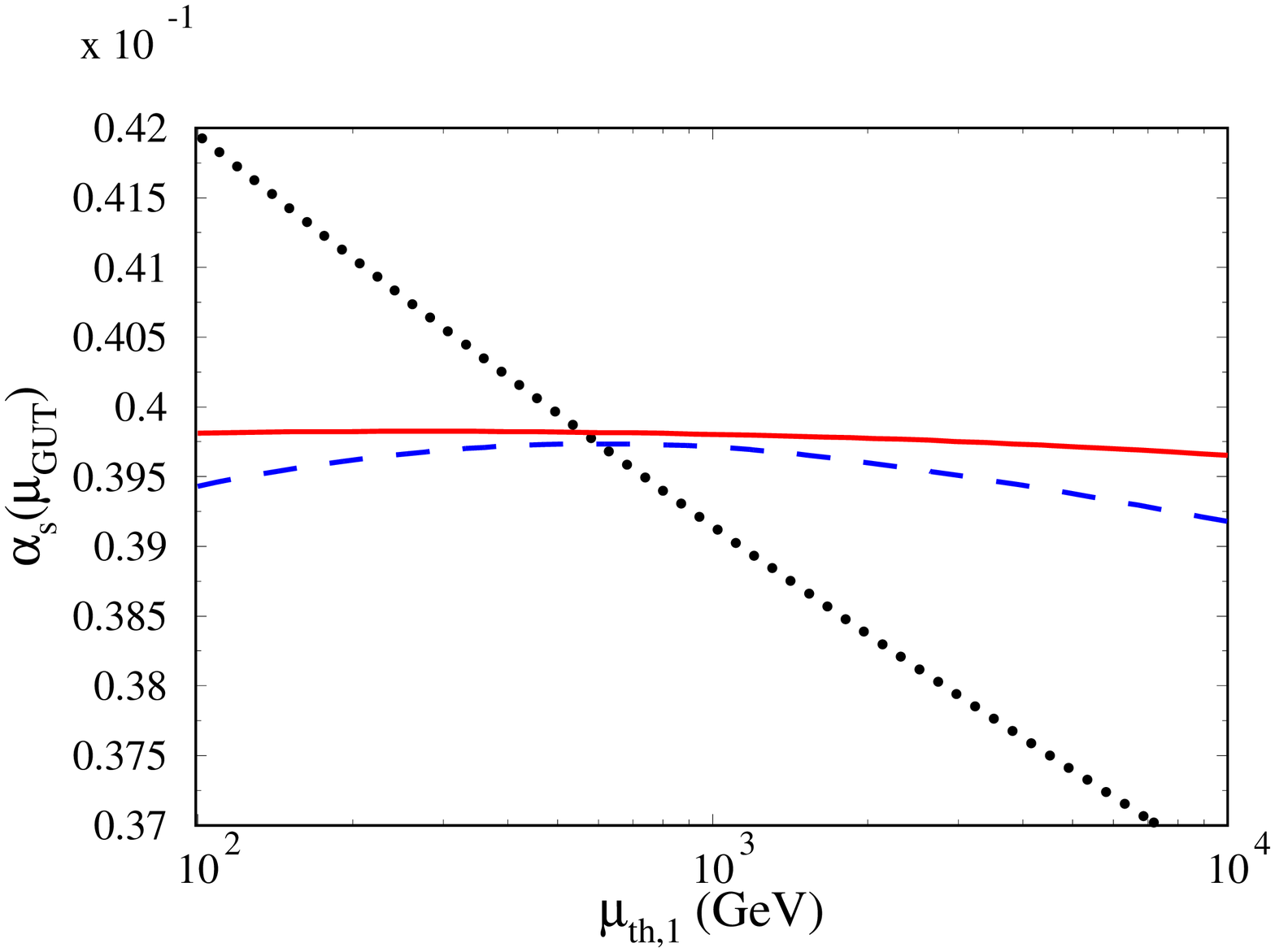,width=18em}
      }
      \\
      \multicolumn{2}{c}{(D)}\\
    \end{tabular}
    \parbox{14.cm}{
      \caption[]{\label{fig::asmuth1}\sloppy
        $\alpha_s(\mu_{\rm GUT})$ as a function 
        of $\mu_{\rm th,1}$ for scenario (B)--(D).
        The dotted, dashed and full lines correspond to
        one-, two- and three-loop running.
        In scenario (D) we set all SUSY masses to $\tilde M$.
        }}
  \end{center}
\end{figure}

\begin{figure}[ht]
  \begin{center}
    \begin{tabular}{cc}
      \epsfig{figure=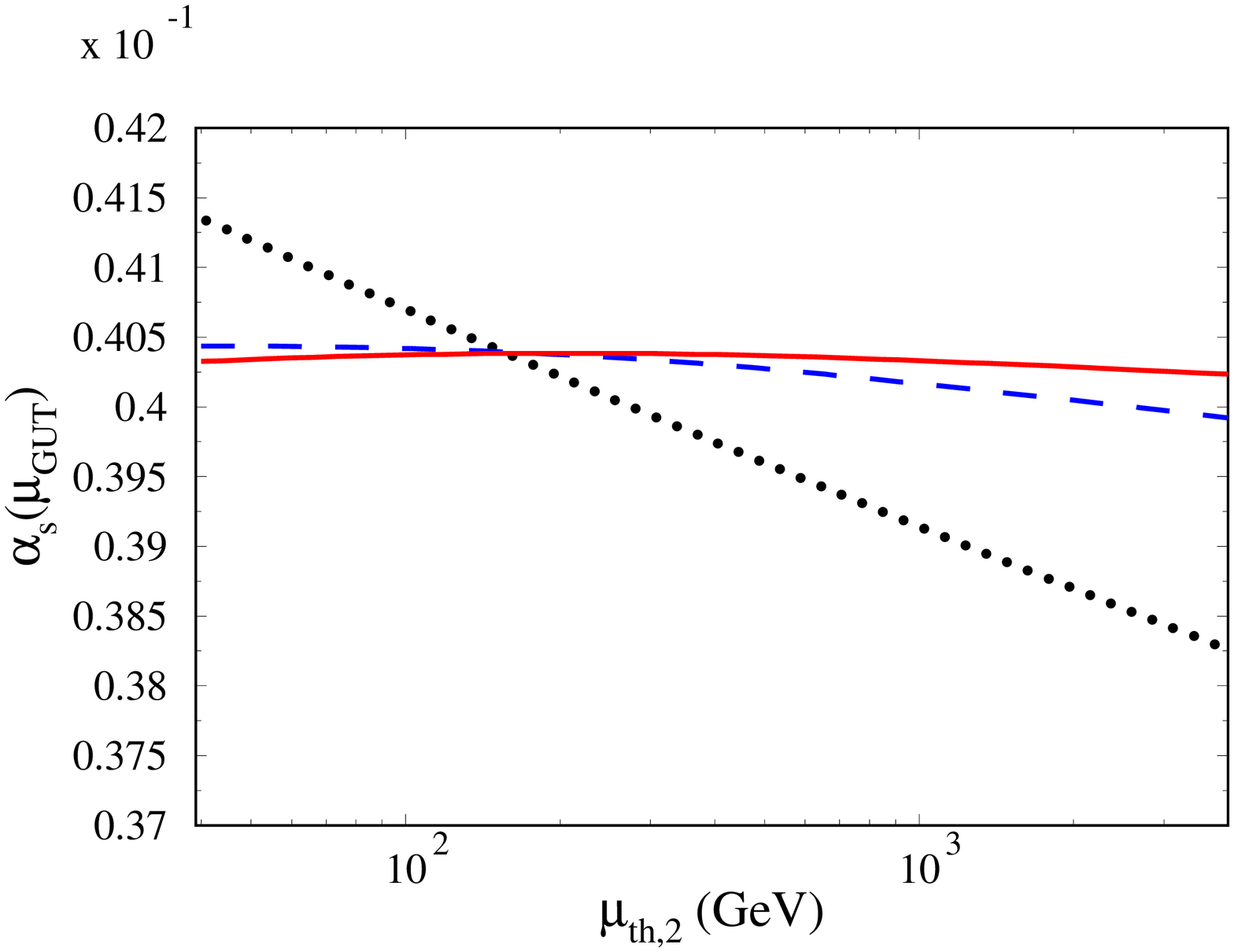,width=18em}
      &
      \epsfig{figure=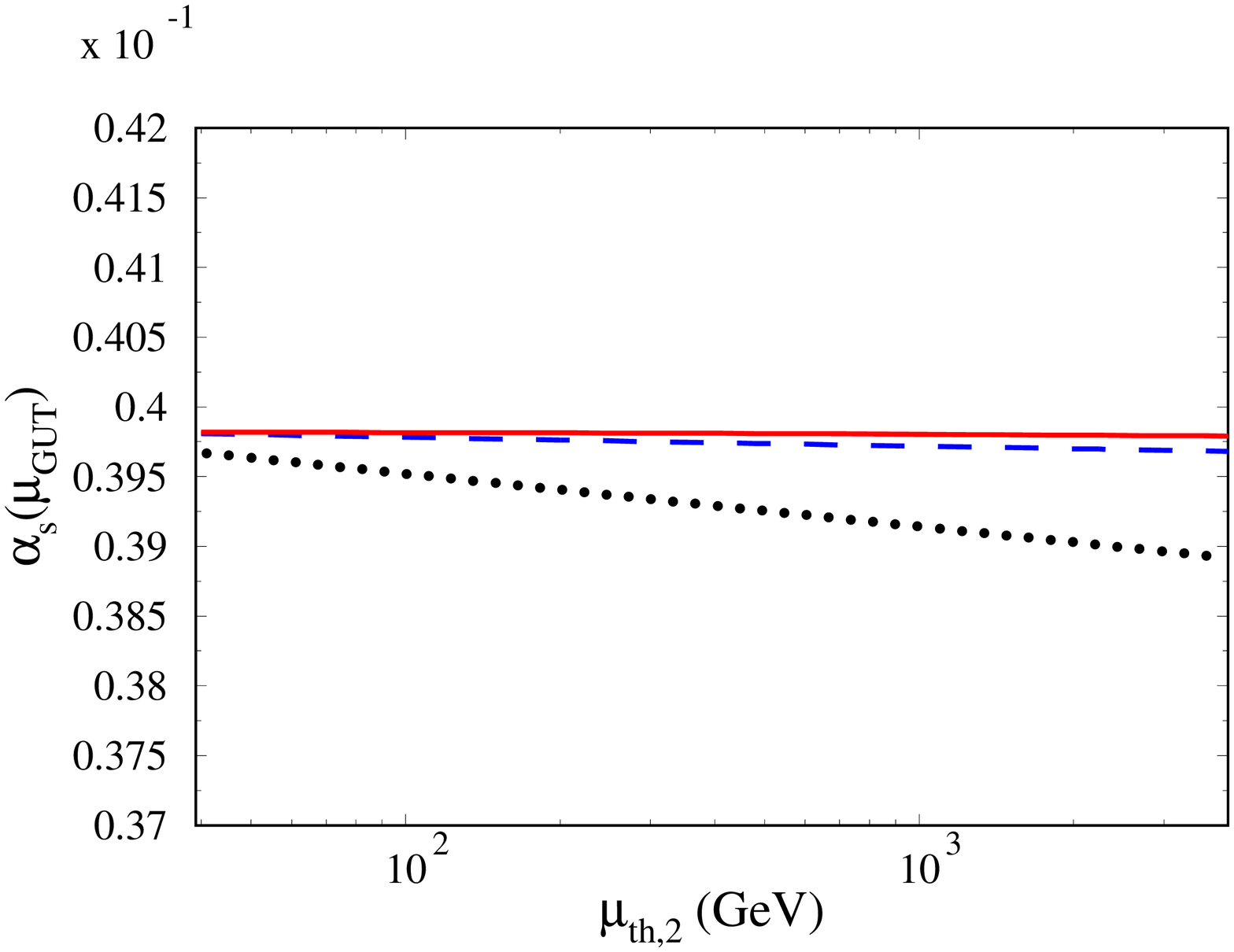,width=18em}
      \\(B) & (C)\\
      \multicolumn{2}{c}{
        \epsfig{figure=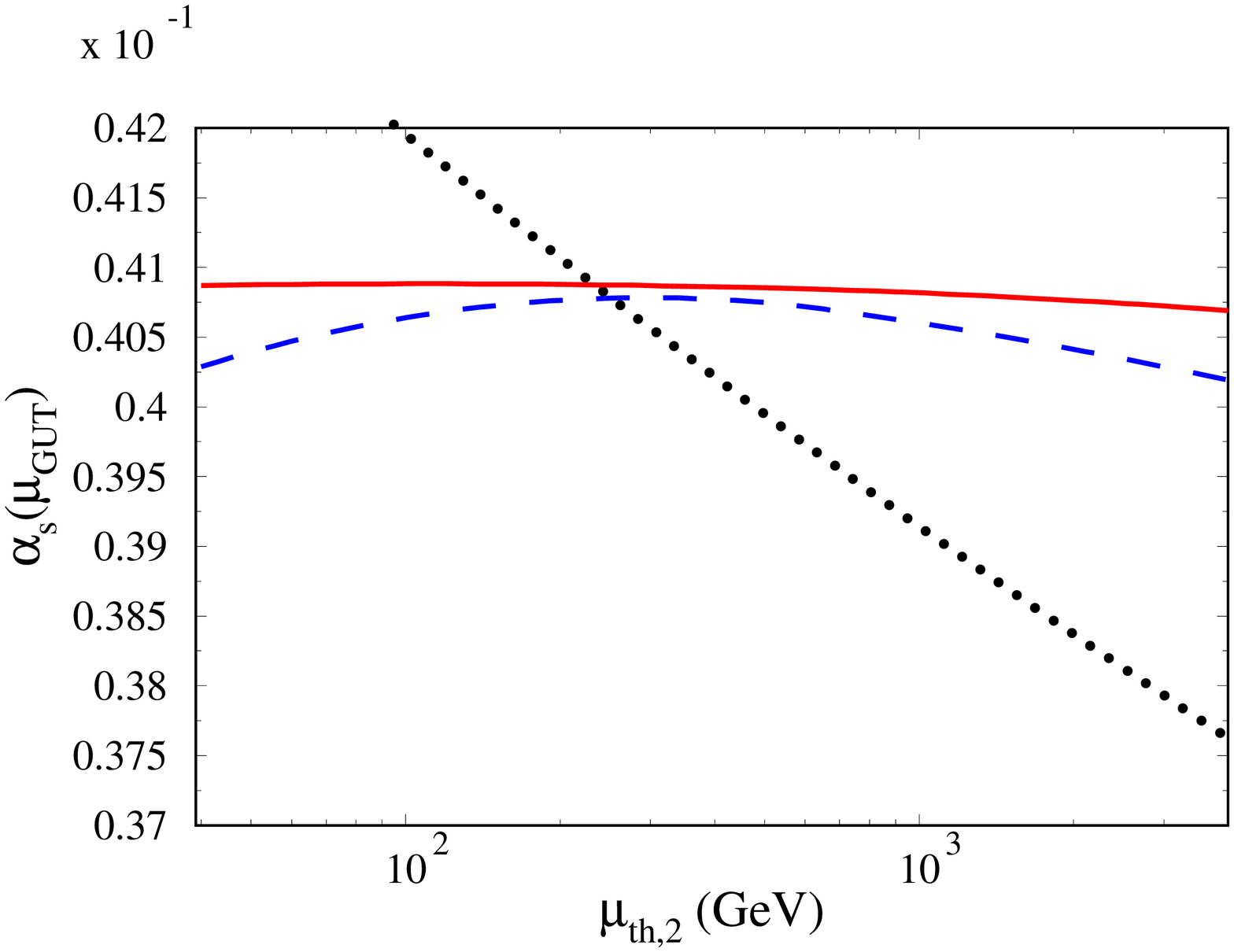,width=18em}
      }
      \\
      \multicolumn{2}{c}{(D)}\\
    \end{tabular}
    \parbox{14.cm}{
      \caption[]{\label{fig::asmuth2}\sloppy
        $\alpha_s(\mu_{\rm GUT})$ as a function 
        of $\mu_{\rm th,2}$ for scenario (B)--(D).
        The dotted, dashed and full lines correspond to
        one-, two- and three-loop running.
        In scenario (D) we set all SUSY masses to $\tilde{m}$.
        }}
  \end{center}
\end{figure}

Figs.~\ref{fig::asmuth1} and~\ref{fig::asmuth2} show the dependence on
one of the scales when the other is kept at its default value.  In
scenario (D), where only one matching scale is present, we set all SUSY
masses to $\tilde M$ when the variation of $\mu_{\rm th,1}$ is
considered (Fig.~\ref{fig::asmuth1}), and to $\tilde m$ when
$\mu_{\rm th,2}$ is varied (Fig.~\ref{fig::asmuth2}).  A drastic
improvement can be observed after including two-loop running and
one-loop matching.  When the three-loop running is included,
the scale variation becomes almost completely flat.

\begin{figure}[ht]
  \begin{center}
    \begin{tabular}{cc}
      \epsfig{figure=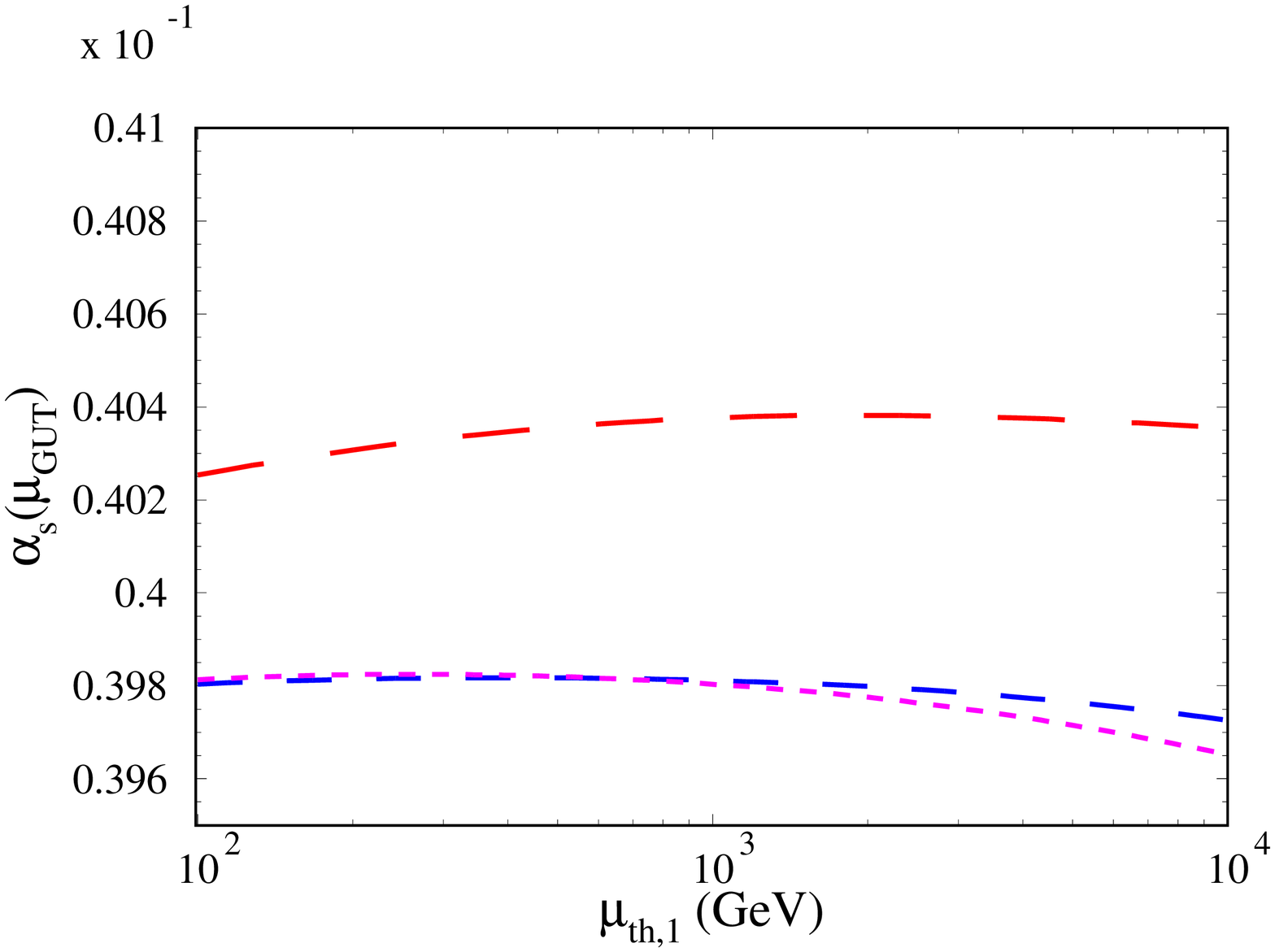,width=18em}
      &
      \epsfig{figure=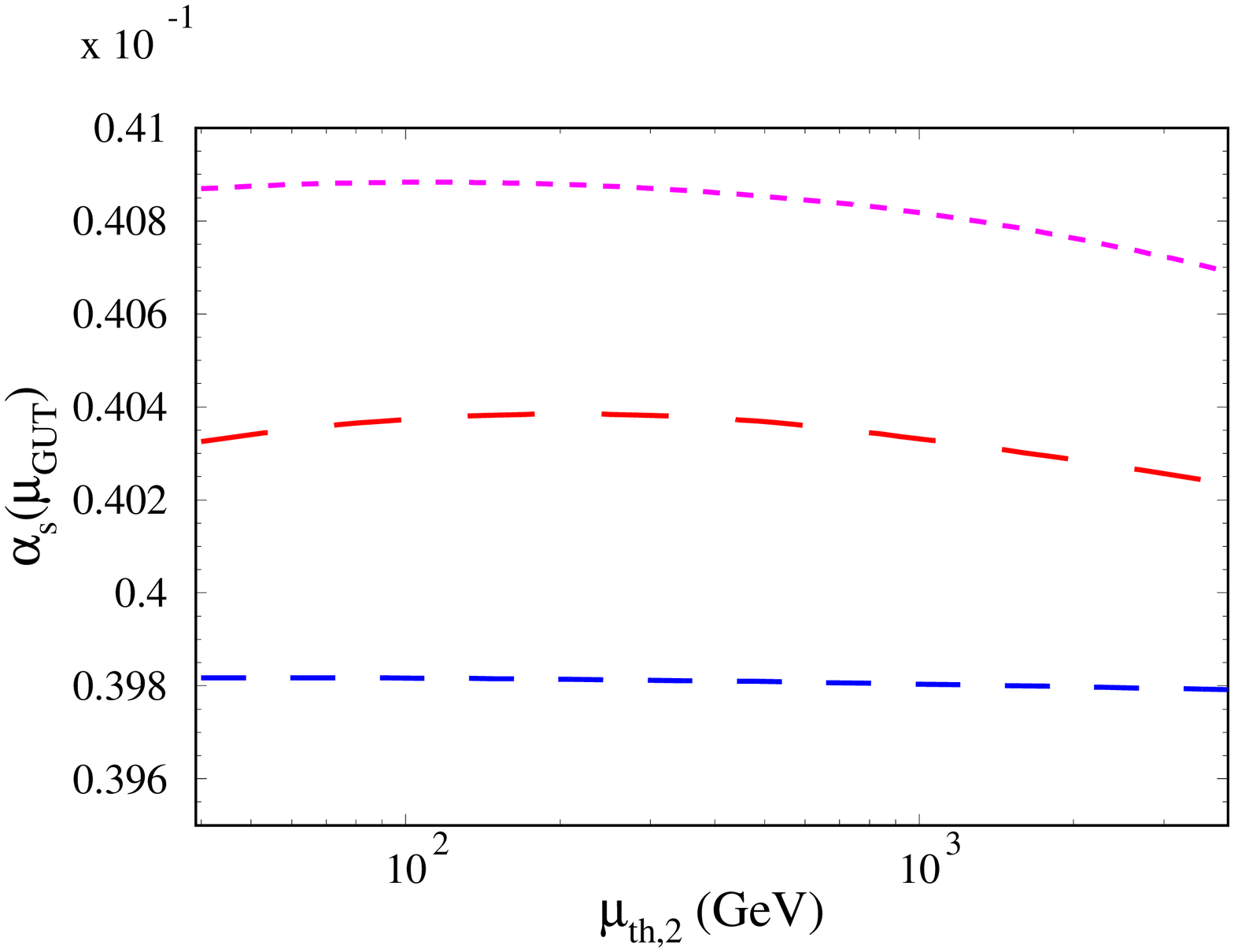,width=18em}
      \\(a) & (b)
    \end{tabular}
    \parbox{14.cm}{
      \caption[]{\label{fig::asth12}\sloppy
        Comparison of $\alpha_s(\mu_{\rm GUT})$ as a function 
        of $\mu_{\rm th,1}$ (a) and $\mu_{\rm th,2}$ (b) where
        for the cases (B), (C) and (D) (from long to short dashes)
        the three-loop results
        are plotted.
        }}
  \end{center}
\end{figure}

In order to study the effect of different mass configurations,
Figs.~\ref{fig::asth12}(a) and (b) compare the three-loop curves of
Figs.~\ref{fig::asmuth1} and~\ref{fig::asmuth2} among scenarios (B)--(D)
(long to short dashes).  One observes differences of the order 0.005
which is about a factor 10 larger than the uncertainty originating from
the current experimental uncertainty in $\alpha_s(M_Z)$.  Estimating the
theoretical uncertainty by the difference between the two- and
three-loop result, we obtain
\begin{eqnarray}
  \alpha_s(\mu_{\rm GUT}) = 0.0398 
  \pm 0.00023 \Big|_{\delta\alpha_s(M_Z)}
  \pm 0.0025   \Big|_{\rm masses}
  \pm 0.00007 \Big|_{\rm th}
  \,.
  \label{eq::asGUT}
\end{eqnarray}

\begin{figure}[ht]
  \begin{center}
    \begin{tabular}{c}
      \epsfig{figure=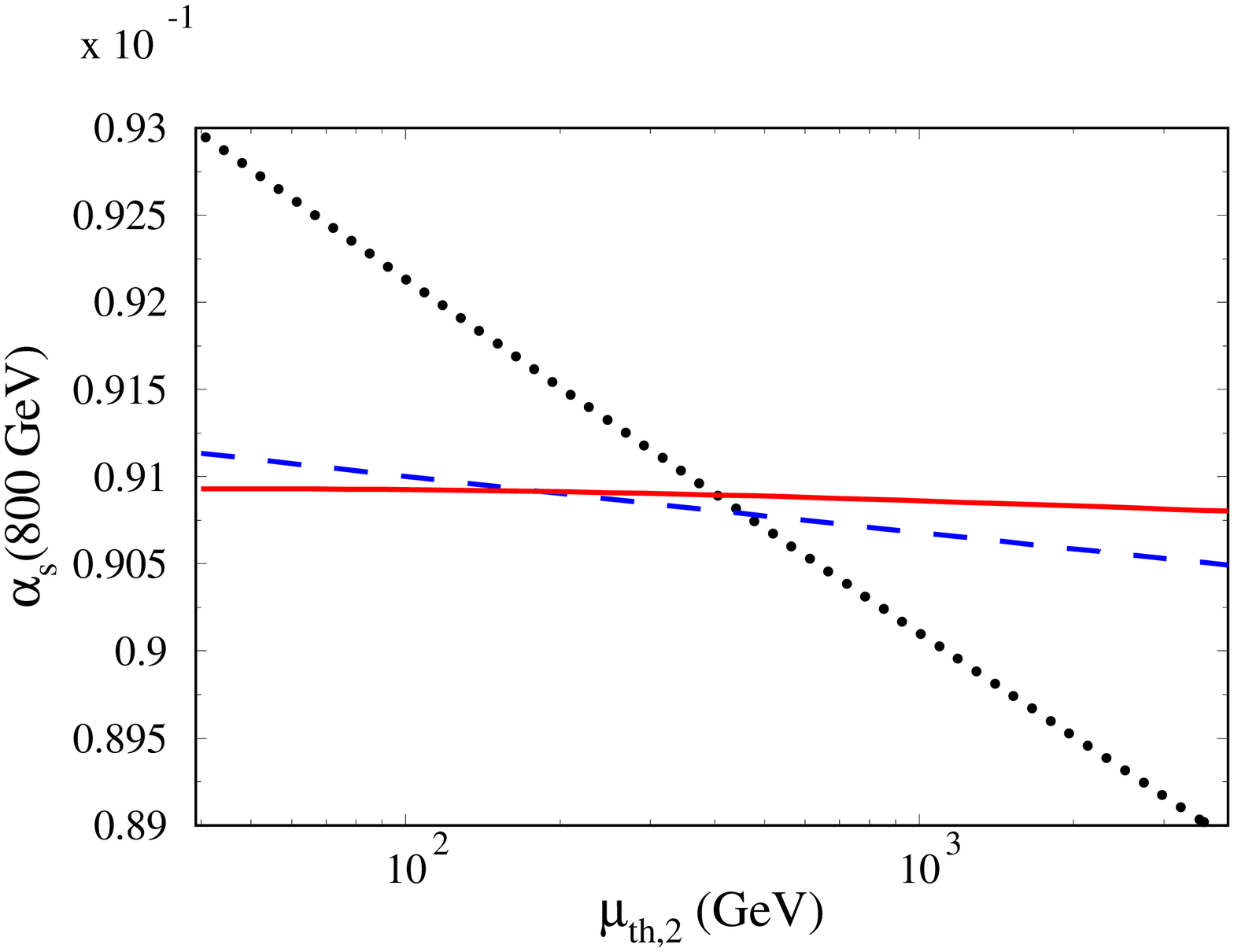,width=36em}
    \end{tabular}
    \parbox{14.cm}{
      \caption[]{\label{fig::asilc}\sloppy
        $\alpha_s(800~\mbox{GeV})$ as a function 
        of $\mu_{\rm th,1}$ for scenario (C).
        The dotted, dashed and full lines correspond to
        one-, two- and three-loop running.
        }}
  \end{center}
\end{figure}

Another practical application of our results is shown in
Fig.~\ref{fig::asilc}, where the strong coupling at a proposed
center-of-mass energy $\sqrt{s}=800$~GeV of a future International
Linear Collider (ILC) is computed.  While for $\mu_{\rm th,2}\approx 200$~GeV
the two- and three-loop result practically give the same value for
$\alpha_s$, for $\mu_{\rm th,2}=2000$~GeV the shift induced by the new
three-loop terms is comparable with the uncertainty induced by the
experimental value for $\alpha_s(M_Z)$
(cf. Tab.~\ref{tab::alphas}). This underlines once more the importance
of the two-loop decoupling and three-loop running terms.

\begin{figure}[ht]
  \begin{center}
    \begin{tabular}{cc}
      \epsfig{figure=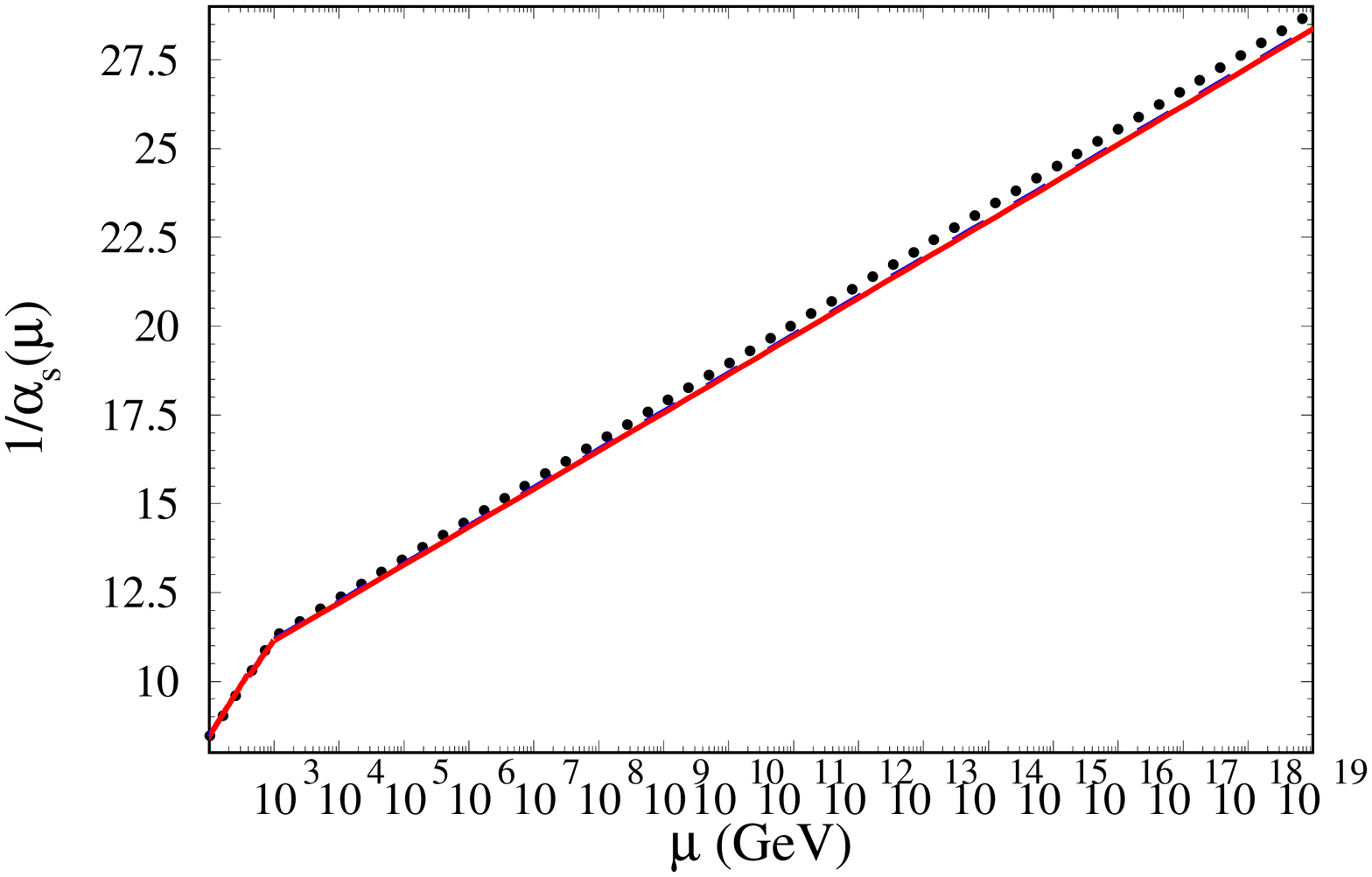,width=18em}
      &
      \epsfig{figure=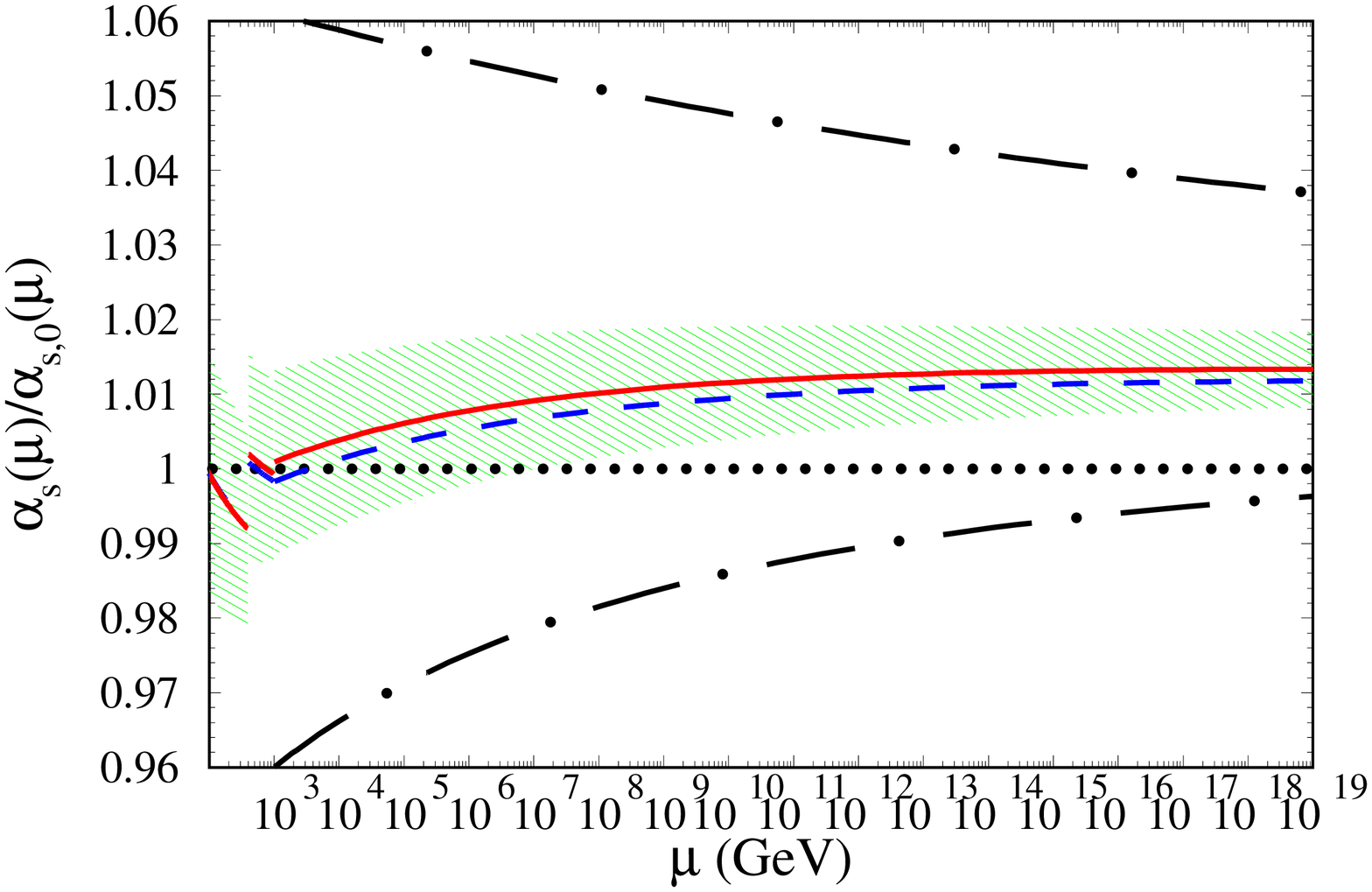,width=18em}
      \\(a) & (b)
    \end{tabular}
    \parbox{14.cm}{
      \caption[]{\label{fig::asmua}\sloppy
        (a) $\alpha_s(\mu)$ as a function 
        of $\mu$ for scenario (C).
        The dotted, dashed and full lines correspond to
        one-, two- and three-loop running.
        (b) Ratio of $\alpha_s(\mu)$ and the corresponding leading order
        value corresponding to $\asMSbar(M_Z)=0.1187$
        for scenario (C). The dotted and dashed lines correspond 
        to one- and two-loop results, respectively. The (green) band 
        corresponds to the uncertainty of $\alpha_s$ as given in 
        Eq.~(\ref{eq::asDRin}).
        The dash-dotted lines corresponds to the three-loop
        result for scenario (D) where the SUSY masses are set to
        400~GeV (upper line) and 2~TeV (lower line).}}
  \end{center}
\end{figure}

Fig.~\ref{fig::asmua}(a) shows $1/\alpha_s$ as a function of $\mu$ from
the weak to the Planck scale, $\mu_{\rm PL}=10^{19}$~GeV, where the
dotted, dashed and solid lines correspond to one-, two- and three-loop
running.  It can be seen that there are kinks at 400~GeV and 1000~GeV
where the slope of the curves changes. A closer look shows that the two-
and three-loop curves actually are discontinuous as $\alpha_s$ jumps due
to the one- and two-loop decoupling relations.  In this plot the
uncertainties from $\alpha_s(M_Z)$ or the various mass patterns of the
supersymmetric particles are hardly visible.  For this reason, we show
in Fig.~\ref{fig::asmua}(b) the ratio
$\alpha_s(\mu)$/$\alpha_{s,0}(\mu)$, where $\alpha_{s,0}(\mu)$ is the
leading order result for scenario~(C) with $\alpha_{s,0}(M_Z)=0.120$.
The scale $\mu$ is varied between $M_Z$ and $\mu_{\rm PL}$. The dotted,
dashed, and solid line correspond to the one-, two-, and three-loop
result of scenario (C), respectively, Note that
below $\mu_{\rm th,2}=\tilde{m}=400$~GeV the difference between the
dashed and full curve is very small. The (green) band
indicates the variation originating from the experimental uncertainty of
$\asDRbar(M_Z)$ (cf. Eq.~(\ref{eq::asDRin})). The dash-dotted lines
correspond to the three-loop results of scenario (D) where the masses
are set to $2\tilde{M}$ (lower) and $\tilde{m}$ (upper),
respectively. They thus reflect the dependence on the mass parameters of
the MSSM.  The corresponding curve for scenario (B) would lie
between the two dash-dotted lines and are therefore not shown.

\begin{figure}[ht]
  \begin{center}
    \begin{tabular}{c}
      \epsfig{bb=95 485 310 720,figure=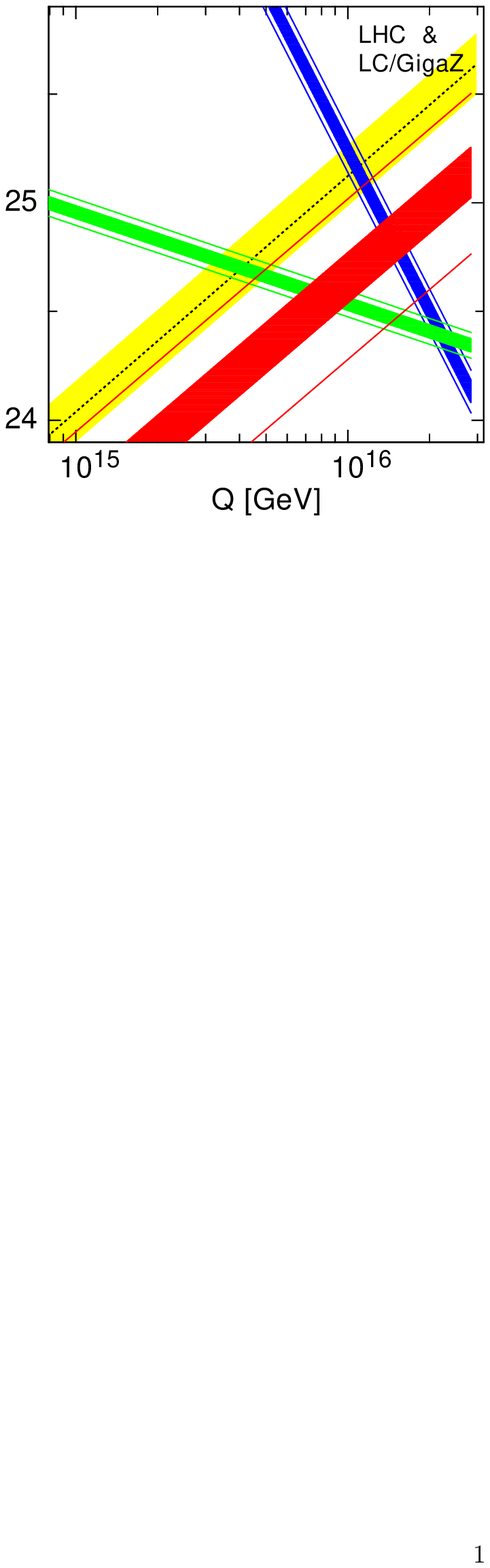,width=32em}
    \end{tabular}
    \parbox{14.cm}{
      \caption[]{\label{fig::couplings}\sloppy
        Running of the inverse couplings $1/\alpha_i$, $i=1$ (blue band),
        $2$ (green band), and $s$ (red band) with two-loop accuracy around
        the unification point defined as the meeting point of $\alpha_1$
        and $\alpha_2$. 
        The wide error bands take into account present data and the
        supersymmetric mass spectrum from LHC measurements within
        mSUGRA. The narrow bands envisage the improvements expected from
        the LHC+LC coherent analysis~\cite{Allanach:2004ud}. The yellow band
        shows our results for $1/\alpha_s$ with three-loop accuracy.           
    }}
  \end{center}
\end{figure}

In Fig.~\ref{fig::couplings} we combine our three-loop accuracy
findings for the evolution of the inverse strong coupling  
at the GUT  scale (yellow band) with the  two-loop order results
for the strong ($\alpha_s$, red band), the electro-magnetic ($\alpha_1$,
blue band) and the weak
coupling ($\alpha_2$, green band) 
(for a precise definition, see 
Ref.~\cite{Allanach:2004ud}, and references therein). 
The wide error bands
correspond to the present experimental accuracy for the gauge couplings
and the supersymmetric mass spectrum of the SPS1a reference point.
The inner bands account for the  expected improvements in 
the absolute errors due to the future experimental analysis at LHC+LC.

\section{\label{sec::concl}Conclusions}

In order to get insight into the mechanism behind Grand Unification, it
is necessary to have precise relations among the various couplings
evaluated at the electro-weak and the GUT scale within the expected
theoretical framework, in our case the MSSM.  In this paper we
provided the two-loop decoupling coefficients which, in combination with
the known three-loop renormalization group equation, establish the
three-loop relation between the strong coupling at $M_Z$ and $\mu_{\rm
GUT}$.  Furthermore, they ensure the independence of the decoupling
procedure from the scale at which the heavy particles are integrated
out, up to higher order terms in $\alpha_s$.

A large variation of $\alpha_s(\mu_{\rm GUT})$ results from the
uncertainty in $\alpha_s(M_Z)$. However, as can be seen from
Eq.~(\ref{eq::asGUT}) and Fig.~\ref{fig::asmua}(b), there is also a
significant uncertainty from the values of the individual mass
parameters. The effect of including the two-loop decoupling relation 
is a significant stabilisation of the dependence of 
$\alpha_s(\mu_{\rm GUT})$ on the matching scales.

The analysis of this paper is based on certain approximations
concerning the mass pattern (cf. Eq.~(\ref{eq::ABCD}) )
within the MSSM spectrum. This results in compact formulae. 
It is desireable to derive in a future calculation more general
formulae and investigate further the dependence on the quark masses.

\vspace*{1em}

\noindent
{\bf Acknowledgements}\\
This work was supported by the Sonderforschungsbereich Transregio 9.  
R.H. is supported by {\it Deutsche Forschungsgemeinschaft} (contract
HA~2990/2-1, Emmy-Noether program). We thank S. Heinemeyer and W. Porod
 for discussions and  useful comments.

\end{document}

%% file: figs/diagrams.tex
\FADiagram{}
\FAProp(0.,10.)(6.,10.)(0.,){/Cycles}{0}
\FALabel(3.,8.93)[t]{$g$}
\FAProp(20.,10.)(14.,10.)(0.,){/Cycles}{0}
\FALabel(17.,11.07)[b]{$g$}
\FAProp(6.,10.)(14.,10.)(0.8,){/Cycles}{0}
\FALabel(10.,5.73)[t]{$g$}
\FAProp(6.,10.)(14.,10.)(-0.8,){/Cycles}{0}
\FALabel(10.,14.27)[b]{$g$}
\FAVert(6.,10.){0}
\FAVert(14.,10.){0}

\FADiagram{}
\FAProp(0.,10.)(6.,10.)(0.,){/Cycles}{0}
\FALabel(3.,8.93)[t]{$g$}
\FAProp(20.,10.)(14.,10.)(0.,){/Cycles}{0}
\FALabel(17.,5.43)[b]{$g$}
\FAProp(6.,10.)(14.,10.)(0.8,){/ScalarDash}{-1}
\FALabel(10.,5.73)[t]{$\tilde t_1$}
\FAProp(6.,10.)(14.,10.)(-0.8,){/ScalarDash}{1}
\FALabel(10.,14.27)[b]{$\tilde t_1$}
\FAVert(6.,10.){0}
\FAVert(14.,10.){0}

\FADiagram{}
\FAProp(0.,10.)(6.,10.)(0.,){/GhostDash}{1}
\FALabel(3.,8.93)[t]{$c$}
\FAProp(20.,10.)(14.,10.)(0.,){/GhostDash}{-1}
\FALabel(17.,6.83)[b]{$c$}
\FAProp(6.,10.)(14.,10.)(0.8,){/GhostDash}{1}
\FALabel(10.,5.73)[t]{$c$}
\FAProp(6.,10.)(14.,10.)(-0.8,){/Cycles}{0}
\FALabel(10.,14.27)[b]{$g$}
\FAVert(6.,10.){0}
\FAVert(14.,10.){0}

\FADiagram{}
\FAProp(0.,10.)(6.6,10.)(0.,){/Cycles}{0}
\FALabel(3.3,8.93)[t]{$g$}
\FAProp(20.,15.)(14.,14.)(0.,){/GhostDash}{-1}
\FALabel(16.7452,15.5489)[b]{$c$}
\FAProp(20.,5.)(14.,6.)(0.,){/GhostDash}{1}
\FALabel(16.7452,4.45109)[t]{$c$}
\FAProp(6.6,10.)(14.,14.)(0.,){/Cycles}{0}
\FALabel(9.51988,13.8512)[br]{$g$}
\FAProp(6.6,10.)(14.,6.)(0.,){/Cycles}{0}
\FALabel(9.85274,6.76457)[tr]{$g$}
\FAProp(14.,14.)(14.,6.)(0.,){/GhostDash}{-1}
\FALabel(15.274,10.)[l]{$c$}
\FAVert(6.6,10.){0}
\FAVert(14.,14.){0}
\FAVert(14.,6.){0}

\FADiagram{}
\FAProp(0.,10.)(6.6,10.)(0.,){/Cycles}{0}
\FALabel(3.3,8.93)[t]{$g$}
\FAProp(20.,15.)(14.,14.)(0.,){/GhostDash}{-1}
\FALabel(16.7452,15.5489)[b]{$c$}
\FAProp(20.,5.)(14.,6.)(0.,){/GhostDash}{1}
\FALabel(16.7452,4.45109)[t]{$c$}
\FAProp(6.6,10.)(14.,14.)(0.,){/GhostDash}{1}
\FALabel(9.85274,13.2354)[br]{$c$}
\FAProp(6.6,10.)(14.,6.)(0.,){/GhostDash}{-1}
\FALabel(9.85274,6.76457)[tr]{$c$}
\FAProp(14.,14.)(14.,6.)(0.,){/Cycles}{0}
\FALabel(15.974,10.)[l]{$g$}
\FAVert(6.6,10.){0}
\FAVert(14.,14.){0}
\FAVert(14.,6.){0}

\FADiagram{}
\FAProp(0.,10.)(6.,10.)(0.,){/Cycles}{0}
\FALabel(3.,8.93)[t]{$g$}
\FAProp(20.,10.)(14.,10.)(0.,){/Cycles}{0}
\FALabel(17.,8.23)[t]{$g$}
\FAProp(6.,10.)(14.,10.)(1.,){/Cycles}{0}
\FALabel(10.,4.93)[t]{$g$}
\FAProp(8.,13.5)(6.,10.)(0.315846,){/Cycles}{0}
\FALabel(5.58149,12.3549)[br]{$g$}
\FAProp(12.,13.5)(8.,13.5)(0.8,){/Cycles}{0}
\FALabel(10.,16.17)[b]{$g$}
\FAProp(12.,13.5)(8.,13.5)(-0.8,){/Cycles}{0}
\FALabel(10.,10.83)[t]{$g$}
\FAProp(12.,13.5)(14.,10.)(-0.310119,){/Cycles}{0}
\FALabel(14.4085,12.3491)[bl]{$g$}
\FAVert(12.,13.5){0}
\FAVert(8.,13.5){0}
\FAVert(6.,10.){0}
\FAVert(14.,10.){0}

\FADiagram{}
\FAProp(0.,10.)(6.,10.)(0.,){/Cycles}{0}
\FALabel(3.,8.93)[t]{$g$}
\FAProp(20.,10.)(14.,10.)(0.,){/Cycles}{0}
\FALabel(17.,5.83)[b]{$g$}
\FAProp(10.,6.)(6.,10.)(-0.434885,){/Straight}{-1}
\FALabel(6.51421,6.51421)[tr]{$t$}
\FAProp(10.,6.)(14.,10.)(0.412689,){/ScalarDash}{1}
\FALabel(13.4414,6.55861)[tl]{$\tilde t_2$}
\FAProp(10.,14.)(10.,6.)(0.,){/Cycles}{0}
\FAProp(10.2,14.)(10.2,6.)(0.,){/Straight}{0}
\FALabel(11.2,10.)[l]{$\tilde g$}
\FAProp(10.,14.)(6.,10.)(0.425735,){/Straight}{1}
\FALabel(6.53252,13.4675)[br]{$t$}
\FAProp(10.,14.)(14.,10.)(-0.412689,){/ScalarDash}{-1}
\FALabel(13.4414,13.0414)[bl]{$\tilde t_2$}
\FAVert(10.,14.){0}
\FAVert(10.,6.){0}
\FAVert(6.,10.){0}
\FAVert(14.,10.){0}

\FADiagram{}
\FAProp(0.,10.)(6.,10.)(0.,){/GhostDash}{1}
\FALabel(3.,8.93)[t]{$c$}
\FAProp(20.,10.)(14.,10.)(0.,){/GhostDash}{-1}
\FALabel(17.,8.93)[t]{$c$}
\FAProp(6.,10.)(14.,10.)(1.,){/GhostDash}{1}
\FALabel(10.,4.93)[t]{$c$}
\FAProp(8.,13.5)(6.,10.)(0.315846,){/Cycles}{0}
\FALabel(5.58149,12.3549)[br]{$g$}
\FAProp(12.,13.5)(8.,13.5)(0.8,){/ScalarDash}{-1}
\FALabel(10.,16.17)[b]{$\tilde t_1$}
\FAProp(12.,13.5)(8.,13.5)(-0.8,){/ScalarDash}{1}
\FALabel(10.,10.83)[t]{$\tilde t_1$}
\FAProp(12.,13.5)(14.,10.)(-0.310119,){/Cycles}{0}
\FALabel(14.4085,12.3491)[bl]{$g$}
\FAVert(12.,13.5){0}
\FAVert(8.,13.5){0}
\FAVert(6.,10.){0}
\FAVert(14.,10.){0}

\FADiagram{}
\FAProp(0.,10.)(5.,10.)(0.,){/Cycles}{0}
\FALabel(2.5,8.93)[t]{$g$}
\FAProp(20.,16.)(15.,16.)(0.,){/GhostDash}{-1}
\FALabel(17.5,14.93)[t]{$c$}
\FAProp(20.,4.)(15.,4.)(0.,){/GhostDash}{1}
\FALabel(17.5,2.93)[t]{$c$}
\FAProp(10.,4.)(5.,10.)(0.,){/Cycles}{0}
\FALabel(6.2515,6.03959)[tr]{$g$}
\FAProp(10.,4.)(15.,4.)(0.,){/Cycles}{0}
\FALabel(12.5,2.93)[t]{$g$}
\FAProp(10.,16.)(10.,4.)(0.,){/Cycles}{0}
\FALabel(4.75,14.)[l]{$g$}
\FAProp(10.,16.)(5.,10.)(0.,){/Cycles}{0}
\FALabel(11.,12.0396)[tl]{$g$}
\FAProp(10.,16.)(15.,16.)(0.,){/Cycles}{0}
\FALabel(12.5,20.6)[t]{$g$}
\FAProp(15.,16.)(15.,4.)(0.,){/GhostDash}{-1}
\FALabel(16.07,10.)[l]{$c$}
\FAVert(10.,16.){0}
\FAVert(10.,4.){0}
\FAVert(5.,10.){0}
\FAVert(15.,16.){0}
\FAVert(15.,4.){0}

\FADiagram{}
\FAProp(0.,10.)(5.,10.)(0.,){/Cycles}{0}
\FALabel(2.5,8.93)[t]{$g$}
\FAProp(20.,16.)(15.,14.)(0.,){/GhostDash}{-1}
\FALabel(18.0757,18.0409)[t]{$c$}
\FAProp(20.,4.)(15.,6.)(0.,){/GhostDash}{1}
\FALabel(16.9243,4.04086)[t]{$c$}
\FAProp(5.,10.)(15.,14.)(0.,){/GhostDash}{1}
\FALabel(10.5757,15.0409)[t]{$c$}
\FAProp(5.,10.)(15.,6.)(0.,){/GhostDash}{-1}
\FALabel(9.42434,7.04086)[t]{$c$}
\FAProp(15.,11.5)(15.,14.)(0.,){/Cycles}{0}
\FAProp(15.,8.5)(15.,11.5)(0.8,){/ScalarDash}{-1}
\FALabel(17.27,10.)[l]{$\tilde t_2$}
\FAProp(15.,8.5)(15.,11.5)(-0.8,){/ScalarDash}{1}
\FALabel(12.73,10.)[r]{$\tilde t_2$}
\FAProp(15.,8.5)(15.,6.)(0.,){/Cycles}{0}
\FAVert(15.,8.5){0}
\FAVert(15.,11.5){0}
\FAVert(5.,10.){0}
\FAVert(15.,14.){0}
\FAVert(15.,6.){0}